\documentclass[runningheads]{llncs}

\usepackage[T1]{fontenc}
\usepackage{rotating}
\usepackage{hyperref}
\usepackage{graphicx}
\usepackage{subcaption}
\newcommand{\key}{\mathsf{key}}
\newcommand{\FK}{\mathcal{F}_K}
\usepackage{tabularx}
\usepackage{makecell}
\usepackage{xcolor}
\newcommand{\rel}{\mathcal{R}}

\newcommand{\col}{\mathsf{col}}
\usepackage{amsmath}
\usepackage{amssymb}
\DeclareMathOperator{\Lex}{Lex}
\usepackage{tcolorbox}
\tcbuselibrary{breakable}
\usepackage{caption}
\usepackage{enumitem}
\usepackage{adjustbox}
\usepackage{multirow}

\usepackage{float}
\usepackage{algorithm}
\usepackage{algpseudocode}
\usepackage{booktabs}   % for \toprule, \midrule, \bottomrule
\usepackage{multirow}   % for \multirow
\usepackage{pifont}     % for \cmark, \xmark
\newcommand{\cmark}{\ding{51}}
\newcommand{\xmark}{\ding{55}}
\begin{document}

\newcommand{\steffen}[1]{\textcolor{red}{Steffen: #1}}

\title{Retrieval-Augmented Generation of
Ontologies from Relational Databases\thanks{ \textbf{Conceptualization:} M.N., N.F., R.B.T., S.S.; 
\textbf{Data curation:} H.-M.T., A.S., M.N., A.A.Y., N.F.; 
\textbf{Formal analysis:} M.N., N.F., S.S.; 
\textbf{Investigation:} M.N., N.F.; 
\textbf{Methodology:} M.N., N.F., R.B.T., S.S.; 
\textbf{Software:} N.F., A.A.Y.; 
\textbf{Validation:} N.F.; 
\textbf{Visualization:} M.N., N.F; 
\textbf{Writing -- original draft:} M.N., A.A.Y., N.F., S.S.; 
\textbf{Writing -- review \& editing:} N.F., S.S. 
}
}

\author{Nadeen Fathallah\inst{1}%\thanks{These authors contributed equally to this work.} 
\and 
Mojtaba Nayyeri\inst{1} \and
Athish A Yogi\inst{1} \and
Ratan Bahadur Thapa\inst{1} \and
Hans-Michael Tautenhahn\inst{3} \and
Anton Schnurpel\inst{3} \and
Steffen Staab\inst{1,2}}
% %
\authorrunning{N. Fathallah et al.}
\institute{Institute for Artificial Intelligence, University of Stuttgart, Stuttgart, Germany \and
University of Southampton, Southampton, UK \and
Universitätsklinikum Leipzig, Leipzig, Germany \\
\email{nadeen.fathallah@ki.uni-stuttgart.de}}
\maketitle          

\begin{abstract}
Deriving OWL ontologies from relational database schemas supports semantic interoperability and downstream tasks such as knowledge graph population, ontology-based data access, graph-based learning, and automated reasoning.
Existing approaches either require substantial expert effort or produce shallow ontologies that reflect the logical schema structure but fail to fully capture domain semantics. We present \textbf{RIGOR} (Retrieval-augmented Iterative Generation of RDB Ontologies), an LLM-driven pipeline that converts relational schemas into semantically rich OWL2DL ontologies with minimal human intervention. For each relational table, \textbf{RIGOR} generates a direct mapping to guarantee schema coverage, then enriches it via retrieval from three sources: relational schema context and documentation, external domain ontologies, and an ontology that grows incrementally as each validated fragment is integrated. A \textit{Gen-LLM} produces provenance-annotated ontology fragments (\emph{delta ontologies}), which are validated and, when needed, corrected by an independent \textit{Judge-LLM} before integration. Guided by foreign-key constraints, the process iterates over relational tables until the full schema is covered. Experiments across three databases spanning two domains show that \textbf{RIGOR} consistently outperforms baseline methods across standard quality metrics while requiring no human oversight.
\end{abstract}
%\input{ISWC26/1-introduction}
%\input{ISWC26/2-related_work}
%\input{ISWC26/3-methodology}
%\input{ISWC26/4-experiments}
%\steffen{I guess that not only ontologies are generated but also ontology-RelDB mappings, not? If yes: Is this input-output relationship reflected everywhere?}
\section{Introduction}

Relational databases~\cite{passant2012bringingsurvey} provide a formally grounded framework for storing, querying, and manipulating data with strong guarantees of consistency and integrity, and serve as the backbone of enterprise applications in healthcare, finance, e-commerce, and government.
However, relational database schemata make it difficult to pose semantic queries~\cite{passant2012bringingsurvey} over their content or to integrate data across heterogeneous sources.

Defining an OWL ontology over a relational database unlocks capabilities that the relational schema alone cannot provide: ontology-based data access (OBDA), which enables conceptually powerful queries over the live database without materialization~\cite{xiao2020virtual}; knowledge graph population, which materializes the data as RDF instances for graph-based learning~\cite{robinson2024relbench} and reasoning~\cite{galkintowards,wu2020comprehensive}; and semantic integration across heterogeneous sources~\cite{wache2001ontology}.
A shallow ontology that mirrors the logical schema already provides some of these benefits; however, a semantically rich ontology---with class hierarchies, disjointness axioms, expressive property restrictions, and alignment to established domain vocabularies---enhances reasoning and cross-system interoperability.

In the past, constructing such an ontology from a database schema was labor-intensive.
Existing automated approaches rely solely on structural cues---table and column names, datatypes, and foreign-key constraints---and produce ontologies that lack the expressive axioms required for effective use by humans and machine reasoners~\cite{ben2021novel}.

LLMs~\cite{jiang2026survey} excel on knowledge-intensive tasks, including comprehending text, encoding conceptual knowledge~\cite{xiong2025tokens}, performing algorithmic reasoning~\cite{zelikman2023parsel}, and generating structured outputs~\cite{liu2024llms,liu2024we}. They have been applied to ontology generation from text~\cite{LLMs4OLbabaei2023llms4ol,qiang2023agentllmonto,llmontomihindukulasooriya2023text2kgbench,ontopopllmontonorouzi2025ontology,lippolis2025ontology}, and recent work applies them to ontology construction from relational schemas~\cite{xiao2025llm4vkg} and to benchmarking such systems~\cite{laskowski2025burr}. However, existing LLM-based approaches target unstructured text and do not address challenges specific to relational schemas, such as interpreting foreign-key semantics, maintaining coherence across interdependent tables, and grounding generation within schema constraints, leading to ontologies that may omit schema elements or misrepresent structural relationships.

\begin{figure*}
\includegraphics[width=13cm]{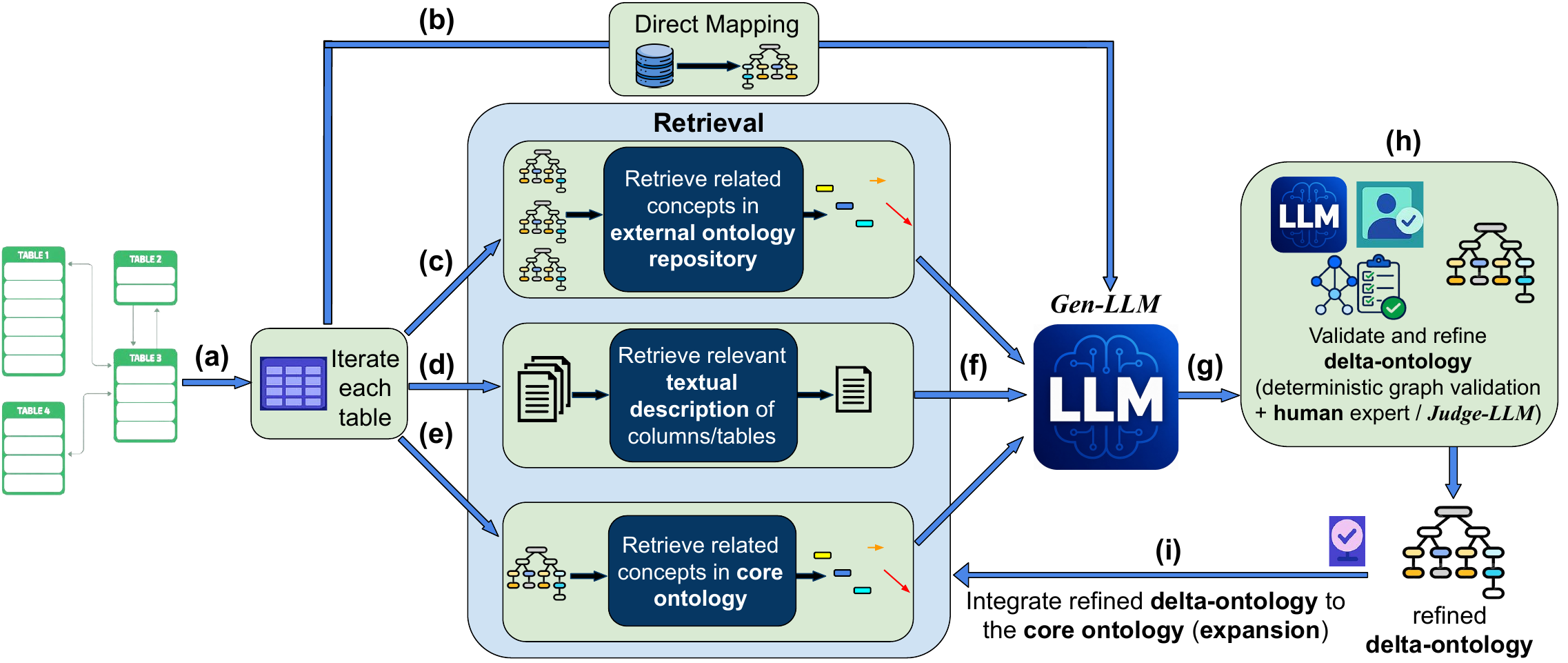}
\caption{Overview of the \textbf{RIGOR} pipeline. Tables are processed iteratively in foreign-key traversal order~(a). For each table, a direct mapping is generated~(b), relevant context is retrieved from an external ontology repository~(c), textual documentation~(d), and the expanding \emph{core ontology}~(e). The direct mapping and retrieved context are passed to the \textit{Gen-LLM} during prompt construction~(f), which produces a semantically enriched \emph{delta ontology}~(g). The \emph{delta ontology} is validated and, if necessary, refined by either a human expert or a \textit{Judge-LLM}, followed by deterministic graph validation~(h), and the accepted result is integrated into the \emph{core ontology} for the following iterations~(i).}
\label{fig1}
\vspace{-1.5em}
\end{figure*}

We introduce \textbf{RIGOR}, an iterative Retrieval-Augmented Generation (RAG) pipeline that uses LLMs to convert relational database schemas into semantically rich OWL ontologies (see Figure~\ref{fig1}). The pipeline processes tables one by one, following foreign-key relationships and grounding each generation step in the relational schema structure. For each table, a deterministic direct mapping provides schema coverage before any LLM involvement. Relevant context is then retrieved from three complementary sources: the relational schema and its textual documentation, the \emph{core ontology} built from previously processed tables, and external domain ontologies. A generative language model (\textit{Gen-LLM}) enriches the direct mapping with labels, class hierarchies, disjointness axioms, and provenance annotations, producing a \emph{delta ontology}. Before integration, an independent \textit{Judge-LLM} validates and, where necessary, corrects the fragment. The pipeline iterates until full schema coverage is achieved, producing an ontology that preserves the schema-to-ontology correspondence needed for both KG population and OBDA. We evaluate \textbf{RIGOR} across two domains (two medical databases and one music-domain database) against W3C Direct Mapping~\cite{directmapsequeda2012directly}, an LLM baseline~\cite{mateiu2023ontology}, and a single-pass non-iterative generation method.
Ontology quality is assessed via consistency checking, semantic similarity, modeling pitfall detection, and expert judgment.
%An ablation study isolates the contribution of each RAG component by systematically enabling or disabling each source and measuring the impact on class, object property, and data property coverage.
\textbf{RIGOR} consistently outperforms all competing methods.

%The \textbf{Chinook} database models a digital media store and contains 11 tables, 67 columns, and 11 foreign-key constraints. It is publicly available and tests cross-domain generalisability beyond the medical setting.

\section{Related Work}

\subsection{Ontology Learning from Relational Databases}

The conversion of relational databases to semantic representations has been studied extensively~\cite{DBLP:journals/ws/VolzHSSS04,astrova2004reverse}.
W3C Direct Mapping and R2RML~\cite{directmapsequeda2012directly,dirmap1arenas2012direct,world2012r2rml} translate each table row into an RDF resource, producing a knowledge graph that preserves the logical schema but does not introduce class hierarchies, property restrictions, or links to domain vocabularies.
BootOX~\cite{jimenez2015bootox} goes further by mapping tables to classes and foreign keys to object properties, then aligning the result with a domain ontology via LogMap; however, it requires a pre-existing target ontology and cannot generate one from scratch.
Tools such as Karma~\cite{karmaknoblock2012semi}, IncMap~\cite{incmappinkel2017incmap}, MIRROR~\cite{de2015mirror}, and D2RQ~\cite{d2rqvergoulis2014data} have been evaluated in the RODI benchmark~\cite{rodipinkel2017rodi}, which showed that all systems struggle with complex schema--ontology mismatches: tables encoding implicit hierarchies, junction tables representing n-ary relationships, and denormalized designs are consistently mis-mapped or left unmapped. Ben et al.~\cite{ben2021novel} learn ontologies from databases using heuristics over foreign keys and data patterns, but their method relies on fixed mapping rules that cannot generalize beyond the patterns they encode, and assumes normalized data; a condition rarely met in real-world clinical or enterprise databases where implicit hierarchies, and cryptic column naming (e.g., \texttt{BM\_DEP} for bone marrow depression) are common. Existing approaches share two limitations: they rely on fixed structural rules that cannot capture implicit semantics, and they produce ontologies without annotations, provenance, or axiomatic depth beyond what the schema directly encodes.

\paragraph{LLM-based ontology learning.}
LLMs have been applied to ontology generation, primarily from unstructured text~\cite{LLMs4OLbabaei2023llms4ol,ontokgenabolhasani2024leveraging,lippolis2025ontology,ontopopllmontonorouzi2025ontology,DBLP:conf/esws/FathallahDGPHK24,fathallah2026extended,DBLP:conf/ekaw/FathallahSA24}.
Babaei et al.~\cite{LLMs4OLbabaei2023llms4ol} evaluated zero-shot LLMs on term typing, taxonomy discovery, and relation extraction; performance was limited, particularly on taxonomy and relation tasks where the models lacked domain grounding.
OntoKGen~\cite{ontokgenabolhasani2024leveraging} improved results using GPT-4 with iterative expert-in-the-loop prompting, but remains dependent on human guidance at each step and targets technical documents rather than structured schemas.
These text-based methods do not address challenges specific to relational schemas, such as resolving junction tables, interpreting foreign-key semantics, or handling denormalized designs.

Xiao et al.~\cite{xiao2025llm4vkg} apply LLMs to virtual knowledge graph construction from database schemas, combining structural pattern recognition with LLM-driven alignment and mapping modules.
Their system assumes a pre-existing target ontology for alignment and operates in a single pass; if the initial generation contains errors, they propagate uncorrected.
That current LLM-based approaches struggle with this task is corroborated by the BURR benchmark~\cite{laskowski2025burr}, which evaluates ontology learning from relational databases using mapping-based metrics: their results show that current LLM-based methods underperform rule-based systems on mapping precision and recall, suggesting that LLMs alone — without structured grounding in the source schema — do not yet produce ontologies of sufficient quality.

\subsection{Foundations} \label{sec:related_foundations}

Our work builds on three established paradigms.

\emph{Retrieval-Augmented Generation}~(RAG)~\cite{raglewis2020retrieval} improves factual accuracy by retrieving relevant context before generation, reducing hallucinations in question answering, dialogue, and code generation~\cite{raggao2023retrieval}; however, RAG alone does not address the structural constraints of ontology engineering, such as type consistency, domain/range correctness, or alignment with OWL profiles.

The \emph{LLM-as-judge} paradigm~\cite{DBLP:journals/corr/abs-2411-15594} uses an auxiliary LLM to evaluate generated outputs; it has been applied to ontology assessment against competency questions~\cite{paulheim2025towards,kommineni2024humanexpertsmachinesllm} and pitfall detection~\cite{lippolis2025ontology}, but only as a post-hoc evaluator, not as an in-loop validator during generation.

\emph{Competency questions}~\cite{gruninger1995methodology} verify whether an ontology captures required domain knowledge; recent work automates their generation using LLMs from ontology triples~\cite{DBLP:conf/esws/RebboudTLT24} and domain text~\cite{DBLP:conf/kgswc/AntiaK23}, but competency question generation from database schemas remains unexplored.

\section{The RIGOR Algorithm}

\subsection{Overview and Design Rationale}

\textbf{RIGOR} processes tables individually for two reasons: it avoids exceeding LLM context-window limits on large schemas, and it allows each generation step to reference classes established for previously processed tables. RIGOR takes as input the schema of a relational database $\mathcal{S}$ (table names, column names with datatypes, and foreign-key constraints), optionally complemented by textual documentation, the pipeline additionally accepts an optional seed core ontology $\mathcal{O}_0$, which may be empty or a pre-existing ontology; starting from an existing ontology allows RIGOR to extend domain knowledge already captured, and outputs an OWL2DL ontology $\mathcal{O}_{\mathrm{final}}$.
%RIGOR operates on the schema and its documentation, not on the data instances stored in the database. This is a deliberate design choice: an ontology derived from data instances (e.g., inferring that column \texttt{HR1} stores names by inspecting cell values) would be tied to the specific dataset and would not transfer to another institution using the same schema with different records. Schema-level construction ensures that the ontology is reusable across deployments. When column names are cryptic or abbreviated, the textual documentation — data dictionaries, DDL comments, or generated descriptions — provides the semantic grounding that column names alone cannot, without requiring access to potentially sensitive patient data.

RIGOR operates on the relational schema and its documentation. When column names are cryptic, textual documentation (data dictionaries, DDL comments) provides the semantic grounding that column names alone cannot; for example, \texttt{BM\_DEP} denotes bone marrow depression.
%not on data instances — ensuring that the resulting ontology is reusable across institutions that share the same schema.
The algorithm iterates table-by-table in foreign-key breadth-first order, starting from root tables, i.e., tables with no foreign-key dependencies ($\FK(r) = \emptyset$). When multiple root tables exist, they are processed in alphabetical order; since root tables have no inter-dependencies, their relative order does not affect the resulting ontology.  For each table, \hyperref[alg:rigor]{RIGOR algorithm} executes the following steps (Figure~\ref{fig1}):
\begin{enumerate}[label=(\alph*),leftmargin=2em,topsep=2pt,itemsep=1pt]
  \item Process tables iteratively in foreign-key breadth-first order.
  
  \item Generate a direct mapping of the table; a deterministic OWL translation.
  
  \item Retrieve related concepts from an external ontology repository.
  
  \item Retrieve textual documentation describing the table and its columns.
  
  \item Retrieve relevant context from the expanding core ontology~$\mathcal{O}_t$.
  
  \item Construct a prompt combining (b--e).
  
  \item Invoke the \textit{Gen-LLM} with the prompt from~(f) to generate a semantically enriched delta ontology~$\Delta\mathcal{O}_r$.
  
 % \item  Generate a semantically enriched delta ontology~$\Delta\mathcal{O}_r$.
  
  \item Validate and, if necessary, correct the delta ontology via a \textit{Judge-LLM}.

  \item Integrate the refined delta ontology into~$\mathcal{O}_t$.
\end{enumerate}

\subsection{Data Structures}
\label{sec:formalization}

We define the data structures used throughout the pipeline; a summary is provided in Appendix~\ref{formalize}, Table~\ref{tab:formalization}.

\paragraph{Relational Database Schema.}
We denote the schema by $\mathcal{S} = (\rel, \key, \col, \FK)$, where
$\mathcal{R}$ is the set of relational tables,
$\key(r)$ returns the primary-key attributes of relation $r \in \mathcal{R}$,
$\col(r)$ returns its set of attributes as \{\emph{$name_{ij}$}, \emph{$datatype_{ij}$}\} pairs, and
$\FK$ is the foreign-key constraint function defined below.

\noindent Let $\mathcal{A} = \bigcup_{r \in \rel} \col(r)$ be the set of all attributes across all relations.

\noindent We define $\FK : \rel \longrightarrow 2^{\mathcal{A} \times \rel \times \mathcal{A}}$ such that $(a, r', a') \in \FK(r)$ indicates that attribute $a$ in relation $r$ references attribute $a'$ in relation $r'$.

\noindent We write ${\FK}_{\text{cols}}(r) = \{a \mid \exists\, r', a'.\; (a, r', a') \in \FK(r)\}$ for the foreign-key attributes of~$r$.

\paragraph{Textual Documentation.}
Let $\mathcal{D}$ be the corpus of natural-language descriptions available for the schema (e.g., DDL comments, a data dictionary, or README files; see Section~\ref{db}).
We define two retrieval functions:
$\textit{docTable} : \rel \to \mathcal{D}$, returning the description of a relation,
and $\textit{docAttr} : \{(r,a) \mid r \in \rel,\; a \in \col(r)\} \to \mathcal{D}$, returning the description of an attribute.

\paragraph{Ontology Representation.}
We adopt OWL2DL as our target language~\cite{owl2dl}. We serialize ontology fragments in OWL~2 Manchester Syntax~\cite{owl2-manchester}: its keyword-based, sentence-like format is more readable for expert review than Turtle or RDF/XML, and its natural-language-like structure is generated more reliably by LLMs~\cite{ni2024l2ceval,jaldi2025knowledge}.

At iteration $t$, the core ontology is
$\mathcal{O}_t = (\mathcal{C}_t, \mathcal{P}_t, \mathcal{A}_t, \mathcal{M}_t)$, where
$\mathcal{C}_t$ is the set of class identifiers,
$\mathcal{P}_t = \mathcal{P}_t^{\mathrm{obj}} \cup \mathcal{P}_t^{\mathrm{data}}$ is the set of property identifiers (object and data properties),
$\mathcal{A}_t$ is the set of OWL2DL axioms, and
$\mathcal{M}_t$ is the set of annotation assertions (\texttt{rdfs:label}, \texttt{rdfs:comment}, provenance metadata).

\paragraph{External Ontology Repository.}
We use a set of $n$ external ontologies \\ $\mathcal{E} = \{\mathcal{O}_1^{E}, \ldots, \mathcal{O}_n^{E}\}$, where each $\mathcal{O}_i^{E} = (\mathcal{C}_i^{E}, \mathcal{P}_i^{E}, \mathcal{A}_i^{E}, \mathcal{M}_i^{E})$, to provide domain context during generation (Section~\ref{external}).
\paragraph{FK-Guided Traversal.}
$\Call{FK-BFS}{\mathcal{S}}$ traverses the foreign-key dependency graph breadth-first.
Root tables ($\FK(r) = \emptyset$) form the initial frontier; child tables are enqueued once all referenced parents have been processed. Ties at any level are broken alphabetically, a deterministic choice that does not affect the resulting ontology since tied tables share no inter-dependencies.
%\paragraph{Direct Mapping.} For each relation $r \in \mathcal{R}$, we apply the W3C Direct Mapping procedure~\cite{directmapsequeda2012directly} to produce an OWL fragment $\mathcal{DM}_r$: the table name becomes a class, each non-FK attribute becomes a data property, and each foreign-key constraint becomes an object property pointing to the class of the referenced table. Before this translation, a deterministic preprocessing step $\text{clean}(r)$ applies two fixed rules: (i)~columns whose names match a predefined list of system-generated ETL artifacts (e.g., \texttt{s\_GUID}, \texttt{load\_timestamp}) are removed, and (ii)~columns whose declared SQL type conflicts with their documented semantics (e.g., a date column typed as \texttt{FLOAT}) are re-typed to the correct XSD datatype. The resulting $\mathcal{DM}_r$ guarantees that every column in the cleaned schema is represented.

\paragraph{Direct Mapping.}
For each relation $r \in \mathcal{R}$, a deterministic preprocessing step $\text{clean}(r)$ removes columns matching a predefined list of system-generated ETL artifacts (e.g., \texttt{s\_GUID}, \texttt{load\_timestamp}) and corrects datatype mismatches where the column name implies a temporal type but the declared SQL type is numeric or string (e.g., a column named \texttt{admission\_date} typed as \texttt{FLOAT} is re-typed to \texttt{xsd:date}).
We then apply the W3C Direct Mapping procedure~\cite{directmapsequeda2012directly} to produce an OWL fragment $\mathcal{DM}_r$: the relational table becomes a class, each non-FK attribute becomes a data property with a universal restriction to its XSD range, and each foreign-key constraint becomes an object property pointing to the referenced class. Primary-key, unique, and non-nullable columns receive existential restrictions. The resulting $\mathcal{DM}_r$ guarantees that every column in the cleaned schema is represented.
%; the \textit{Gen-LLM} then enriches this fragment semantically rather than generating from scratch.
%; the \textit{Gen-LLM} then enriches this OWL fragment semantically rather than generating from scratch.
%removes ETL-artifact columns (e.g., system-generated fields such as \texttt{s\_GUID} with no domain meaning) and corrects datatype mismatches (e.g., date columns erroneously typed as \texttt{FLOAT}).

\paragraph{Delta Ontology.}
For each relation $r \in \mathcal{R}$, the \textit{Gen-LLM} produces a delta ontology $\Delta \mathcal{O}_r = (\mathcal{C}_r, \mathcal{P}_r, \mathcal{A}_r, \mathcal{M}_r)$ by enriching $\mathcal{DM}_r$ using retrieved context from the documentation~$\mathcal{D}$, the core ontology~$\mathcal{O}_t$, and external ontologies~$\mathcal{E}$.
Delta ontologies are iteratively integrated into the core ontology (Section~\ref{sec:integration}).

\paragraph{Target Output.}
After all $T = |\mathcal{R}|$ iterations, the output is $\mathcal{O}_{\mathrm{final}} = \mathcal{O}_T$, which must satisfy:
\begin{enumerate}[label=\alph*),leftmargin=1.8em,topsep=0pt]
  \item \textbf{Schema Coverage}: every table is represented by one or more classes.
  \item \textbf{Semantic Alignment}: attributes and foreign-key constraints are modeled as data properties or object properties, respectively.
  \item \textbf{Provenance}: every ontology element is annotated with a \texttt{prov:wasDerivedFrom} assertion referencing its source table or column.
  \item \textbf{External Alignment}: where applicable, elements are linked to external ontologies via \texttt{owl:equivalentClass} or \texttt{rdfs:subClassOf}.
\end{enumerate}

\paragraph{Lexical Views.}
We define lexical views to extract the human-readable strings from ontologies and schemas; these strings serve as input to the embedding-based retrieval described in Section~\ref{sec:dense-retrieval}.

\noindent The \emph{ontology lexical view} collects all label and comment strings:
\[
\Lex(\mathcal{O})
\;=\;
\bigl\{\,
  \lambda \;\bigm|\;
  \langle e,\,p,\,\lambda\rangle\in\mathcal{M},\;
  e \in \mathcal{C}\cup\mathcal{P}^{\mathrm{obj}}\cup\mathcal{P}^{\mathrm{data}},\;
    p \in \{\texttt{rdfs{:}label},\;\texttt{rdfs{:}comment}\}
\bigr\}.
\]

\noindent The \emph{schema lexical view} collects table and attribute names:
\[
  \Lex(\mathcal{S})
  \;:=\;
  \bigl\{\textit{name}(r)\mid r\in\rel\bigr\}
  \;\cup\;
  \bigl\{\textit{name}(a)\mid r\in\rel,\;a\in\col(r)\bigr\}.
\]

\noindent The \emph{universe of texts} at iteration~$t$ combines all lexical views and documentation:
\[
\hspace*{-3em}
 \mathcal{X}_t
  \;:=\;
  \Lex(\mathcal{S})
 \;\cup\;
 \Lex(\mathcal{O}_t)
  \;\cup\;
 \bigcup_{i=1}^{n}\Lex\!\bigl(\mathcal{O}^{E}_i\bigr)
  \;\cup\;
  \text{range}(\textit{docTable})
  \;\cup\;
  \text{range}(\textit{docAttr}).
\]
Every $x \in \mathcal{X}_t$ is a natural-language string suitable for embedding (Section \ref{sec:dense-retrieval}).
$\mathcal{X}_t$ grows with each iteration as validated delta ontologies expand $\mathcal{O}_t$.

\subsection{Embedding-based Retrieval}
\label{sec:dense-retrieval} Passing all available context to the \textit{Gen-LLM} would exceed LLM context-window limits and introduce noise; dense retrieval selects only the most relevant elements from each source, and handles synonymy better than keyword matching.
Each generation step retrieves from three sources --- the
core ontology, the documentation, and the external
repository --- using dense embeddings.

\noindent\textbf{Embedding function.}
We use \texttt{all-MiniLM-L6-v2}~\cite{wang2020minilm}, a distilled SentenceTransformer that achieves consistently strong performance across similarity, retrieval, and clustering tasks on independent benchmarks such as MTEB~\cite{muennighoff2023mteb}, while remaining lightweight enough for per-iteration re-indexing.
We define $\phi : \mathcal{X}_t \to \mathbb{R}^d$ ($d = 384$).
All embeddings are indexed using Faiss~\cite{douze2024faiss} for approximate nearest-neighbor search by cosine similarity:
$\text{sim}(\mathbf{v}_x, \mathbf{v}_y) = \frac{\mathbf{v}_x \cdot \mathbf{v}_y}{\|\mathbf{v}_x\| \, \|\mathbf{v}_y\|}.$

\noindent\textbf{Retrieval.}
For a target relation $r$, we construct a compound query 
$u = \textit{name}(r) \circ \textit{className}(r) \circ 
\textit{colNames}(r) \circ \textit{fkTargets}(r)$, concatenating 
the table name, its class name, its column names, and the names 
of FK-referenced tables. The query embedding is 
$\mathbf{v}_u = \phi(u)$.
We retrieve the top-$k$ most relevant elements from each source, where $\mathrm{proj}_1$ extracts the element from each (element, score) pair:
\begin{align}
\mathcal{N}^{\mathcal{O}_t}_u &=
  \mathrm{proj}_1\Bigl(
    \text{Top}_k\bigl\{\,
      \bigl(x,\,\text{sim}(\mathbf{v}_u,\mathbf{v}_x)\bigr)
      \mid x \in \Lex(\mathcal{O}_t)
    \bigr\}
  \Bigr), \\[2pt]
\mathcal{N}^{\mathcal{E}}_u &=
  \mathrm{proj}_1\Bigl(
    \text{Top}_k\bigl\{\,
      \bigl(x,\,\text{sim}(\mathbf{v}_u,\mathbf{v}_x)\bigr)
      \mid x \in \textstyle\bigcup_{i=1}^{n}\Lex(\mathcal{O}^{E}_i)
    \bigr\}
  \Bigr), \\[2pt]
\mathcal{N}^{\text{Doc}}_u &=
  \mathrm{proj}_1\Bigl(
    \text{Top}_k\bigl\{\,
      \bigl(x,\,\text{sim}(\mathbf{v}_u,\mathbf{v}_x)\bigr)
      \mid x \in \text{range}(\textit{docTable})
        \cup \text{range}(\textit{docAttr})
    \bigr\}
  \Bigr).
\end{align}
We set $k = 3$ for all retrieval sources. This balances context relevance against prompt length: smaller values risk missing related concepts, while larger values introduce noise and consume LLM context-window budget needed for the direct mapping and generation instructions. The three result sets $\mathcal{N}^{\mathcal{O}_t}_u$, $\mathcal{N}^{\mathcal{E}}_u$, and $\mathcal{N}^{\text{Doc}}_u$ (where $u$ is the compound query derived from $r$) form the context for the \textit{Gen-LLM} prompt.

\subsection{Prompt Construction and Generation}
\label{sec:prompt-construction}

The direct mapping provides structural coverage but produces only mechanical translations; semantic enrichment requires context that deterministic rules cannot provide. A single prompt per table combines the direct mapping $\mathcal{DM}_r$ (Section~\ref{sec:formalization}) with the retrieved context $\mathcal{N}^{\mathcal{O}_t}_r$, $\mathcal{N}^{\mathcal{E}}_r$, and $\mathcal{N}^{\text{Doc}}_r$ (Section~\ref{sec:dense-retrieval}), and instructs the \textit{Gen-LLM} to enrich $\mathcal{DM}_r$ with six operations that deterministic translation cannot provide:
\begin{enumerate}[leftmargin=1.8em, topsep=2pt, itemsep=1pt]
    \item \textbf{Annotation:} Add \texttt{rdfs:label} and
    \texttt{rdfs:comment} to every class and property,
    expanding abbreviated names (e.g., \texttt{BM\_DEP}
    $\to$ \texttt{rdfs:label "Bone Marrow Depression"@en})
    and documenting units where applicable.
    \item \textbf{Hierarchy:} Declare \texttt{SubClassOf}
    axioms where a natural specialization exists between
    classes.
    \item \textbf{Value documentation:} For integer columns
    encoding categorical or ordinal concepts, add
    \texttt{rdfs:comment} annotations documenting the value
    meanings (e.g., \textit{``ECOG score ranges from 0
    (fully active) to 5 (dead)''}).
    \item \textbf{Disjointness:} Declare \texttt{DisjointWith} axioms between sibling classes where logically justified.
    \item \textbf{Existential restrictions:} Assert
    \texttt{$p_a$ some $\tau$} for \textsc{not null},
    \textsc{unique}, or primary-key columns.
    \item \textbf{External alignment:} Link classes or
    properties to external ontologies via
    \texttt{owl:equivalentClass} or \texttt{rdfs:subClassOf}
    when supported by retrieved context.
\end{enumerate}

\noindent The prompt additionally imposes structural constraints on the output to reduce modeling errors: type consistency (no URI declared as both object and data property), exactly one \texttt{rdfs:domain} and one \texttt{rdfs:range} per property, standard XSD datatypes for data property ranges, and no self-referential object properties unless explicitly justified. The full prompt template is given in Appendix~\ref{box:prompt-example}, \autoref{fig:promptbox}.

\subsection{Validation and Refinement} \label{sec:validation-refinement} LLM-generated ontology fragments may contain modeling errors that neither the prompt constraints nor the \emph{Gen-LLM's} self-consistency can prevent. Each delta ontology $\Delta \mathcal{O}_r$ undergoes two validation stages before integration into the core ontology (\autoref{fig1}-(h)): a \textit{Judge-LLM} and deterministic graph validation.

\noindent\textbf{Judge-LLM assessment.} The \textit{Judge-LLM} evaluates $\Delta \mathcal{O}_r$ against 14 criteria organized into 3 severity levels, grounded in established ontology quality frameworks~\cite{vrandevcic2009ontology}, the OOPS! pitfall catalogue~\cite{poveda2014oops}, and the OWL2DL specification~\cite{owl2dl}. The full prompt template is provided in Appendix~\ref{box:prompt-example}, \autoref{fig:judgeLLM}. The \textit{Judge-LLM} returns one of three decisions: \textsc{Approved} (accepted as-is), \textsc{Approved\_With\_Corrections} (accepted with a corrected fragment), or \textsc{Rejected} (critical violations require regeneration).

\noindent\textbf{Deterministic graph validation.} After parsing the accepted fragment into an RDF graph, a rule-based validator checks type consistency (no URI declared as both object and data property), domain/range cardinality, XSD-only datatype ranges, and full schema coverage. %This catches structural errors that the \textit{Judge-LLM} may miss.

\noindent\textbf{Retry mechanism.} Failures from either stage are returned to the \textit{Gen-LLM} with the identified issues as correction instructions. This cycle repeats up to $k = 2$ attempts; if still rejected, the \textit{Judge-LLM} selects the best candidate among its corrected fragment and the direct mapping $\mathcal{DM}_r$. The loop is formalized in Algorithm~\ref{alg:rigor}, lines 10--21. We set $k = 2$ following evidence from the LLM self-refinement literature~\cite{madaan2023self,shinn2023reflexion}, which shows that corrective feedback yields diminishing returns beyond two iterations.

%This cycle repeats up to $k = 2$ attempts. If still rejected, the \textit{Judge-LLM}'s corrected version is used if available; otherwise, the last output is accepted with a warning. 

\subsection{Delta Ontology Merging}
\label{sec:integration}
The \textit{Gen-LLM} reuses IRIs from the core ontology for shared classes and properties, so merging reduces to axiom-set union without requiring conflict resolution. Concretely, the accepted delta ontology $\Delta\mathcal{O}_r$ is merged into
$\mathcal{O}_{t-1}$ to form $\mathcal{O}_t$:
$\mathcal{C}_t = \mathcal{C}_{t-1} \cup \mathcal{C}_r$,
$\mathcal{P}_t = \mathcal{P}_{t-1} \cup \mathcal{P}_r$,
$\mathcal{A}_t = \mathcal{A}_{t-1} \cup \mathcal{A}_r$,
$\mathcal{M}_t = \mathcal{M}_{t-1} \cup \mathcal{M}_r$
(\autoref{fig1}-(i); Algorithm~\ref{alg:rigor}, line~22).
The pipeline proceeds to the next table in breadth-first order until all $T = |\mathcal{R}|$ tables have been processed, yielding $\mathcal{O}_{\mathrm{final}}$. 
%Because the \textit{Gen-LLM} reuses IRIs from the core ontology for shared classes and properties, name collisions do not arise; OWL2DL profile compliance is preserved since the union of consistent DL axiom sets over a shared signature remains consistent.

\begin{algorithm}
\caption{\textbf{: RIGOR}}
\label{alg:rigor}
{\fontsize{7}{8}\selectfont
\begin{algorithmic}[1]
\Require Schema $\mathcal{S} = (\mathcal{R}, \key, \col, \FK)$,
         documentation $\mathcal{D}$, external ontologies $\mathcal{E}$,
         seed core ontology $\mathcal{O}_0$, retry budget $k$
\Ensure  Enriched ontology $\mathcal{O}_{\mathrm{final}}$

\State $\langle r_1, \ldots, r_T \rangle \gets
        \Call{FK-BFS}{\mathcal{S}}$
        \Comment{FK-guided traversal}

\For{$t = 1$ \textbf{to} $T$}
    \State $r \gets r_t$
    \State $r' \gets \text{clean}(r)$
           \Comment{Remove ETL cols, fix types}
    \State $\mathcal{DM}_r \gets \Call{DirectMap}{r'}$
           \Comment{Direct Mapping}

    \Statex \Comment{--- Retrieve context ---}
   \State $\mathcal{N}^{\mathcal{O}} \gets
           \textsc{Retrieve}_{k}(\Lex(\mathcal{O}_{t-1}),\; r)$
    \State $\mathcal{N}^{\mathcal{D}} \gets
           \textsc{Retrieve}_{k}(\mathcal{D},\; r)$
    \State $\mathcal{N}^{\mathcal{E}} \gets
           \textsc{Retrieve}_{k}(\Lex(\mathcal{E}),\; r)$

    \Statex \Comment{--- Generate delta ontology ---}
    \State $\Delta\mathcal{O}_r \gets \Call{Gen-LLM}{
           \mathcal{DM}_r,\; \mathcal{N}^{\mathcal{O}},\;
           \mathcal{N}^{\mathcal{D}},\; \mathcal{N}^{\mathcal{E}}}$

    \Statex \Comment{--- Judge-LLM validation loop ---}
    \For{$j = 1$ \textbf{to} $k + 1$}
        \State $(\textit{decision},\; \textit{issues},\;
               \Delta\mathcal{O}'_r) \gets
               \Call{Judge-LLM}{\Delta\mathcal{O}_r,\;
               \mathcal{DM}_r,\; \mathcal{O}_{t-1}}$
        \If{$\textit{decision} = \text{Approved}$}
            \State \textbf{if}
                   $\Call{Validate}{\Delta\mathcal{O}_r,\,
                   \mathcal{DM}_r,\,\mathcal{O}_{t-1}}$
                   \textbf{then break}
        \ElsIf{$\textit{decision} =
               \text{Approved\_With\_Corrections}$}
            \State $\Delta\mathcal{O}_r \gets \Delta\mathcal{O}'_r$
            \State \textbf{if}
                   $\Call{Validate}{\Delta\mathcal{O}_r,\,
                   \mathcal{DM}_r,\,\mathcal{O}_{t-1}}$
                   \textbf{then break}
        \ElsIf{$j \leq k$}
            \State $\Delta\mathcal{O}_r \gets \Call{Gen-LLM}{
                   \mathcal{DM}_r,\; \mathcal{N}^{\mathcal{O}},\;
                   \mathcal{N}^{\mathcal{D}},\;
                   \mathcal{N}^{\mathcal{E}},\; \textit{issues}}$
        \EndIf
    \EndFor

    \State \textbf{if} no \textbf{break} occurred \textbf{then}
           $\Delta\mathcal{O}_r \gets \Call{Judge-LLM-Select}{
           \Delta\mathcal{O}_r,\; \mathcal{DM}_r}$

    \State $\mathcal{O}_t \gets \mathcal{O}_{t-1} \cup
           \Delta\mathcal{O}_r$
           \Comment{Merge into core}
\EndFor

\State \Return $\mathcal{O}_{\mathrm{final}} \gets \mathcal{O}_T$
\end{algorithmic}
}
\end{algorithm}

\section{Experiments}

\subsection{Evaluation Databases}

\subsubsection{Relational Database Schemas} \label{RD}

We evaluate on two medical databases and one music-domain database that differ in origin, size, and schema complexity. The \textbf{Liver Cancer Database} is a liver cancer registry comprising 16 tables, 350 columns, and 15 foreign key constraints. Because this database is not public, it prevents LLM test data leakage~\cite{paulheim2025towards}. The \textbf{eICU-CRD}~\cite{article} (eICU Collaborative Research Database) models intensive care unit admissions (31 tables, 559 columns, 30 foreign-key constraints), available via PhysioNet \cite{goldberger2000physiobank}. The two databases are complementary: the liver cancer registry has a compact, specialist schema with domain-specific encoded values (e.g., ECOG scores), while eICU-CRD has a broad, general ICU schema with many tables and an extensive foreign-key structure. To further assess cross-domain generalisability beyond medicine, we use the \textbf{Chinook} database \cite{chinookdatabase}, which models a digital music store with artists, albums, tracks, playlists, customers, invoices, and employees (11 tables, 67 columns, and 11 foreign-key constraints). Together, they test whether RIGOR generalizes across schema size, complexity, and domain specificity.
%It is publicly available and tests cross-domain generalisability beyond the medical setting.
%The schema is shared in our repository; patient records are not.
\subsection{Competency Question Generation} \label{CQ_GEN}
We use competency questions as a post-hoc evaluation instrument.  We prompt an LLM to generate five competency questions per table from the database schema using chain-of-thought prompting~\cite{DBLP:conf/nips/Wei0SBIXCLZ22}; each is a natural-language question paired with an answer explaining how the ontology would address it. The generated competency questions were verified by a domain expert.
The generation prompt is shown in Appendix \ref{box:prompt-example}, \autoref{fig:cq_gen}, and examples of generated competency questions are shown in Appendix \ref{eval_details}, Table \ref{tab:physicalexam-cqs}.

%\subsubsection{Data for Downstream KG Population} \label{KG_DATA} For the liver cancer database, we use a clinical export containing 7{,}841 patient records across 16 tables, accessible only within the institution. For eICU-CRD, we generated 100 synthetic ICU patients using Synthea\footnote{\url{https://github.com/synthetichealth/synthea}} and mapped them to the eICU-CRD schema; the synthetic dataset preserves schema structure and realistic value distributions while containing no real patient data. For Chinook, we use its instance data, which contains 15{,}607 records across all 11 tables.

%, covering artists, albums, tracks, playlists, customers, invoices, invoice lines, employees, genres, media types, and playlist-track associations.

\subsection{External Ontology Repository} \label{external}

For the biomedical databases, we selected four ontologies from BioPortal\footnote{\label{bioportal}\url{https://bioportal.bioontology.org/}} on the recommendation of domain experts, chosen to cover complementary aspects of the clinical domain represented in our evaluation databases: disease classification (ICD-10), disease semantics (Human Disease Ontology), cell biology (Cell Ontology), and nutrition (Ontology for Nutritional Studies). For Chinook database, we use the Music Ontology \footnote{\label{music_onto} \texttt{https://www.kaggle.com/datasets/danielesantini/wikidata-music-ontology}}, Performed Music Ontology \footnote{\label{performed_music}\texttt{https://github.com/LD4P/PerformedMusicOntology}}, and DOREMUS Ontology\footnote{\label{dorem_onto} \texttt{https://github.com/DOREMUS-ANR/doremus-ontology}}. Statistics are reported in Appendix~\ref{exp_details}, Table~\ref{struct}.

\subsection{Experimental Setup}

\textbf{Database Documentation.} \label{db}
We used GPT-4o to generate natural-language descriptions of database entities, which were reviewed by domain experts.

\noindent\textbf{Large Language Models.} 
\newline
\textit{Generation.}
We used three LLMs from different families, all accessed via the OpenRouter API\footnote{\url{https://openrouter.ai}}:
\texttt{Claude-Opus-4.6} (Anthropic)\footnote{\label{claude}\texttt{anthropic/claude-opus-4-6}},
\texttt{Mistral-Small-24B-Instruct}\footnote{\label{mistral}\texttt{mistralai/mistral-small-24b-instruct-2501}}, and
\texttt{DeepSeek-V3}\footnote{\label{deepseek}\texttt{deepseek/deepseek-chat}}.
Each model was used as the generation model in all three LLM-based methods (Section~\ref{onto_methods}). In RIGOR, the same model additionally served as the \textit{Judge-LLM}.

\noindent\textit{Evaluation.}
Competency questions were generated using \texttt{Mistral-Small-24B-Instruct}\footref{mistral} (Section~\ref{CQ_GEN}).
Evaluation performed by \emph{LLM-as-a-Judge} (Section~\ref{sec:eval_strategies}, Strategy~6) used \texttt{GPT-5.4}\footnote{\url{https://openai.com/index/introducing-gpt-5-4/}} (OpenAI), from an independent model family to eliminate self-evaluation bias.

\noindent\textbf{Computational Resources and Cost.}
Ontology generation ran on an HPC node with two NVIDIA A100 GPUs. Generation runtime, API requests, token usage, and cost by database and model are reported in Appendix~\ref{eval_details}, Table~\ref{tab:generation_cost_runtime}.

\subsection{Ontology Generation Methods} \label{onto_methods}

%We compare four methods across a controlled spectrum, each adding one capability over the previous (a Detailed comparison of the methods is shown in Appendix~\ref{exp_details}, Table~\ref{tab:method_comparison}).
We compare four methods in a controlled spectrum, each adding one capability over the previous one (Appendix~\ref{exp_details}, Table~\ref{tab:method_comparison}).

\noindent\textbf{(i) W3C Direct Mapping}~\cite{directmapsequeda2012directly}: the standard deterministic baseline used throughout the relational-to-ontology literature~\cite{rodipinkel2017rodi}. It translates the schema into OWL without any LLM involvement, establishing the lower bound of what structural translation alone can achieve (Section~\ref{sec:formalization}). Since Sequeda et al.~\cite{directmapsequeda2012directly} define the direct mapping formally rather than providing a reusable implementation, we implemented the baseline according to their mapping specification.

\noindent\textbf{(ii) Baseline}~\cite{mateiu2023ontology}: the LLM-based ontology generation method of Mateiu et al.~\cite{mateiu2023ontology}, originally proposed for ontology generation from natural language, adapted here to database schemas. Since no implementation is provided, we reproduce the method from the paper's prompt and methodological description, replacing the natural-language input with table schemas. The LLM receives only the table schema in a zero-shot setting — no retrieved context, no validation — isolating the contribution of the LLM's parametric knowledge. Full prompt is shown in Appendix \ref{box:prompt-example}, Figure \ref{fig:baseline}.

\noindent\textbf{(iii) Non-Iterative}: extends the baseline (ii) with full schema context, FAISS-retrieved documents, and external-ontology concepts. This represents the current state of practice in RAG-based ontology generation~\cite{jaldi2025knowledge} and isolates the effect of retrieval augmentation from iterative refinement. Full prompt is shown in Appendix \ref{box:prompt-example}, Figure \ref{fig:noniterative_prompt}.

\noindent\textbf{(iv) RIGOR}: the full pipeline, starting from an empty core ontology ($\mathcal{O}_0 = \emptyset$) and adding the direct mapping as a generation seed, the expanding core ontology as cross-table context, and \textit{Judge-LLM} validation with bounded retry (Algorithm~\ref{alg:rigor}). This tests whether iterative enrichment and validation yield gains beyond what retrieval alone provides.

\subsection{Evaluation Strategies} 
\label{sec:eval_strategies}
%A Comprehensive Benchmark for Evaluating LLM-Generated Ontologies
We employ eight complementary evaluation techniques:

\noindent\textbf{1. Syntax Validity Check:}
Each ontology was checked for OWL2DL compliance using RDFLib.

\noindent\textbf{2. Logical Consistency Check:} Logical consistency was verified using the HermiT reasoner \cite{glimm2014hermit}.

\noindent\textbf{3. Modeling Pitfalls:} We used OOPS!~\cite{poveda2014oops} to detect common modeling pitfalls in the generated ontologies, consistent with prior works \cite{lippolis2025ontology}. OOPs! scans for 41 pitfalls (\textit{P01--P41}) and categorizes each finding as \textit{critical}, \textit{important}, or \textit{minor} based on its impact on ontology quality.

\noindent\textbf{4. Structural Analysis:} We report the number of classes, object and data properties, axioms, annotations, and provenance assertions per ontology using Ontometrics \cite{lozano2004ontometric}.

%\noindent\textbf{5. Semantic Coverage:} We measure how well each ontology preserves the source schema by computing semantic matches between schema elements and ontology elements. Specifically, we embed schema table and column names, together with ontology class and property names, using \texttt{all-MiniLM-L6-v2} and compute cosine similarity. A table or column is considered covered if at least one corresponding ontology element exceeds a similarity threshold of 0.55, empirically tuned and consistent with prior work~\cite{galli2024performance}.

\noindent\textbf{5. Semantic Coverage:} To know how much of the source schema is represented in the ontology, we embed schema table and column names, as well as ontology class and property names, using \texttt{all-MiniLM-L6-v2}. A schema table or column is considered covered if at least one ontology element exceeds a cosine-similarity threshold of 0.55, empirically tuned and consistent with prior work~\cite{galli2024performance,DBLP:conf/semtech4stld/FathallahSA25}.

%\noindent\textbf{6. Ontology Quality through Competency Question Performance:} We adopt the \emph{LLM-as-a-Judge} (GPT-5.4) (Section~\ref{sec:related_foundations}) for comparative ontology evaluation. Competency questions (Section~\ref{CQ_GEN}) serve as the evaluation anchor: they define what domain knowledge each table's ontology should capture, giving the \emph{LLM-as-a-Judge} concrete criteria against which to assess ontologies. For each table, we compare four ontology fragments: Direct Mapping \cite{directmapsequeda2012directly} and the three LLM-based methods generated with Claude (Baseline \cite{mateiu2023ontology}, Non-Iterative, and RIGOR); Section~\ref{onto_methods}. Each fragment is extracted as a class-centric chunk containing the matching \texttt{owl:Class} with its properties, annotations, and ranges. The fragments are presented to an \emph{LLM-as-a-Judge}, anonymized as A--D to prevent method-name bias. The \emph{LLM-as-a-Judge} receives the four ontology fragments, the table schema, and 5 competency questions per table (80 for liver cancer, 155 for eICU-CRD, 55 for Chinook), and ranks them 1st--4th while scoring each on 0--5 along seven dimensions grounded in established ontology quality frameworks~\cite{article,vrandevcic2009ontology,tartir2005ontoqa}: accuracy, completeness, conciseness, adaptability, clarity, consistency, and domain enrichment. Scores are averaged across tables. Dimension definitions and the full prompt are in Appendix~\ref{box:prompt-example}, Figure~\ref{fig:llm-As-Judge}.

\noindent\textbf{6. Ontology Quality through Competency Question Performance:}
We use GPT-5.4 as an \emph{LLM-as-a-Judge} (Section~\ref{sec:related_foundations}), with competency questions (Section~\ref{CQ_GEN}) as table-level evaluation anchors specifying the domain knowledge each ontology should capture. For each table, we compare four anonymized class-centric ontology fragments (A--D): Direct Mapping~\cite{directmapsequeda2012directly} and three Claude-generated LLM-based methods (Baseline~\cite{mateiu2023ontology}, Non-Iterative, and RIGOR; Section~\ref{onto_methods}). Each fragment contains the matching \texttt{owl:Class}, its properties, annotations, and ranges. The judge receives the fragments, table schema, and five competency questions per table (80 for liver cancer, 155 for eICU-CRD, 55 for Chinook), then ranks fragments 1st--4th and scores each on seven 0--5 dimensions from established ontology quality frameworks~\cite{article,vrandevcic2009ontology,tartir2005ontoqa}: accuracy, completeness, conciseness, adaptability, clarity, consistency, and domain enrichment. Scores are averaged across tables; scoring quality dimension definitions and the full prompt are in Appendix~\ref{box:prompt-example}, Figure~\ref{fig:llm-As-Judge}.

%\noindent\textbf{7. Expert Evaluation:}
%Ten experts with backgrounds in ontology engineering and knowledge representation (profiles and years of experience in Appendix~\ref{eval_details}, Figure~\ref{fig:expert_profile}) evaluated ontology quality on the two medical databases via a structured online survey. We selected two tables per database that (a)~are present in all four ontologies, ensuring fair comparison, and (b)~cover distinct modeling challenges: \texttt{general\_aftercare} (liver cancer; clinical follow-up with multiple foreign keys) and \texttt{complication} (liver cancer; encoded severity grades and part-whole relationships to procedures); \texttt{hospital} (eICU-CRD; central administrative entity with few columns) and \texttt{patient} (eICU-CRD; densely connected entity referenced by most other tables). For each selected table, experts received the table schema, 5 competency questions, and four anonymized ontology fragments (Direct Mapping \cite{directmapsequeda2012directly} and the three Claude-generated LLM-based ontologies, labeled A--D to prevent method-name bias). They scored each fragment on the same seven dimensions used in Strategy~6 (0--5 scale) and ranked the fragments 1st--4th. Inter-rater agreement on method rankings was measured using Kendall's $W$~\cite{kendall1939problem}, which quantifies concordance among multiple raters on ordinal rankings.

\noindent\textbf{7. Expert Evaluation:}
Ten experts in ontology engineering and knowledge representation (profiles in Appendix~\ref{eval_details}, Figure~\ref{fig:expert_profile}) evaluated ontology quality on the two medical databases via a structured survey over two tables per database. We selected tables that appear in all four ontologies and cover distinct modeling challenges: \texttt{general\_aftercare} (liver cancer; clinical follow-up with multiple foreign keys), \texttt{complication} (liver cancer; encoded severity grades and procedure part-whole relations), \texttt{hospital} (eICU-CRD; compact administrative entity), and \texttt{patient} (eICU-CRD; densely connected entity referenced by most tables). Experts received the table schema, 5 competency questions, and four anonymized ontology fragments (Direct Mapping~\cite{directmapsequeda2012directly} and three Claude-generated LLM-based ontologies, labeled A--D). They scored each on the same seven dimensions as Strategy~6 (0--5 scale) and ranked fragments 1st--4th. Inter-rater agreement was measured using Kendall's~$W$~\cite{kendall1939problem}, which quantifies concordance among multiple raters on ordinal rankings.

\noindent\textbf{8. Ontology-learning Benchmark:}
We use the BURR~\cite{laskowski2025burr} benchmark to compare RIGOR to established ontology-learning systems. BURR measures mapping-based F1 for classes~(C), relations~(R), and attributes~(A) by comparing database-to-ontology mappings with a gold standard. We use BURR's ISWC database, covering the International Semantic Web Conference domain: conferences, papers, persons, and topics. The database has 9 tables, 46 columns, and 11 foreign keys. We generate the RIGOR ontology with Claude, using GPT-5.4 for documentation and SWRC\footnote{\url{https://ontoware.org/swrc/swrc/SWRCOWL/swrc_updated_v0.7.1.owl}} and BIBO\footnote{\url{https://www.dublincore.org/specifications/bibo/bibo/bibo.rdf.xml}} as external ontologies. BURR evaluates \emph{mappings} rather than ontology structure; therefore, we export RIGOR's \texttt{prov:wasDerivedFrom} annotations as a D2RQ mapping~\cite{d2rqvergoulis2014data}.

%, and provided free-text feedback.

%\noindent\textbf{8. Downstream KG Population:} We populate each RIGOR ontology with the data in Section~\ref{KG_DATA}. Rows become RDF individuals, columns become typed property assertions, and foreign-key references become object property links. We report the number of individuals, property assertions, tables populated versus skipped, and column binding rate. Unmatched tables reveal gaps between the ontology and the data.

\subsection{Results} \label{results}

We evaluate 30 ontologies: 3 LLMs $\times$ 3 LLM-based methods
$\times$ 3 databases, plus 3 Direct Mapping \cite{directmapsequeda2012directly} ontologies. Detailed results in Appendix~\ref{eval_details}, Tables~\ref{tab:pitfalls}--\ref{tab:expert_scores}.

\noindent\textbf{Strategies 1--2: Syntax and Consistency.}
All 30 ontologies are valid OWL2DL (RDFLib) and logically
consistent (HermiT).

\noindent\textbf{Strategy 3: Modeling Pitfalls.} OOPS! reports substantially fewer pitfalls for all RIGOR variants (Appendix~\ref{eval_details}, Table~\ref{tab:pitfalls}). On eICU-CRD, Direct Mapping~\cite{directmapsequeda2012directly} produces 665 pitfalls, and every Baseline~\cite{mateiu2023ontology} ontology exceeds 400; RIGOR reduces this to single digits for Claude and Mistral, and to 26 for DeepSeek. On the liver cancer database and Chinook, a similar pattern holds: RIGOR consistently produces one to two orders of magnitude fewer pitfalls than all other methods, with all three RIGOR variants achieving zero pitfalls on Chinook. This reduction is not merely a post-processing effect: while some baseline pitfalls (e.g., missing domain/range declarations, \textit{P11}) can be repaired mechanically, such repairs do not provide meaningful labels, provenance, or class hierarchies.

\noindent\textbf{Strategy 4: Structural Analysis.} RIGOR is the only method that consistently produces annotation assertions, provenance triples, and \texttt{SubClassOf} axioms (Appendix~\ref{eval_details}, Table~\ref{tab:structural_metrics}). Direct Mapping~\cite{directmapsequeda2012directly} and the Baseline method~\cite{mateiu2023ontology} produce none of these; the Non-Iterative method produces annotations in some configurations but never provenance or class hierarchies. RIGOR is also the only method to generate \texttt{owl:disjointWith} axioms (13 on Chinook with Mistral), a task known to be particularly challenging even for dedicated methods~\cite{volker2015automatic}; no other method produced any disjointness axioms.

\noindent\textbf{Strategy 5: Semantic Coverage.} RIGOR maintains high schema coverage (Appendix~\ref{eval_details}, Table~\ref{tab:schema_coverage}) while adding the structures reported in Strategy~4. Across all databases, RIGOR variants cover all tables; column coverage ranges from 0.77 to 1.00 across databases and models. Direct Mapping~\cite{directmapsequeda2012directly} achieves comparable column coverage by construction but without annotations, provenance, or subclass axioms. The Non-Iterative method, despite retrieving the same context, achieves lower column coverage, confirming that single-pass generation omits many schema properties.

\noindent\textbf{Strategies 6 and 7: Competency-Question and
Expert Evaluation.} Table~\ref{tab:eval_combined} reports quality scores and ranking frequencies from two complementary evaluators: an \textit{LLM-as-a-Judge} (GPT-5.4) over all tables and ten human experts over four representative tables from the medical databases. RIGOR ranks first on all three databases under the \textit{LLM-as-a-Judge} and leads in expert rankings on both medical databases. The strongest gains appear on \textit{Clarity} and \textit{Domain Enrichment}, where annotations, value documentation, and class hierarchies are directly rewarded (Appendix~\ref{eval_details}, Tables~\ref{tab:cq_comparative},~\ref{tab:expert_scores}). The Non-Iterative method performs worst on eICU-CRD, where the \textit{LLM-as-a-Judge} ranks it last on all 31 tables, confirming that retrieved context alone, without schema grounding and validation, is insufficient. As shown in Appendix \ref{eval_details}, Table~\ref{tab:kendall_w}, the ten experts showed significant agreement on method rankings for all four evaluated tables, with Kendall's~$W$ ranging from 0.39 (moderate agreement) to 0.72 (strong agreement). RIGOR was ranked first in 28 of 40 expert evaluations (70\%).
%https://github.com/citiususc/rdb2owl-bench

\noindent\textbf{Strategy 8: Ontology-learning Benchmark.} RIGOR achieves the highest class F1~(0.73), relation F1~(0.67), and attribute F1~(0.52) among ontology-generating systems on the ISWC database of BURR~\cite{laskowski2025burr} (Table~\ref{tab:burr}). Laskowski et al.~\cite{laskowski2025burr} concluded that current LLM-based methods underperform rule-based systems on mapping recovery; RIGOR reverses this finding by combining LLM generation with schema-grounded provenance, achieving competitive mapping scores as a byproduct of semantically rich ontology construction.
\begin{table}
%\captionsetup{font=scriptsize}
\centering
\caption{Ontology quality: mean scores (0--5, averaged across seven quality dimensions) and ranking frequency for Direct Mapping~\cite{directmapsequeda2012directly} and the three LLM-generated ontologies. \textit{LLM-as-a-Judge} (GPT-5.4) over all parsed tables. Experts: 10 human experts over four representative tables from the two medical databases. Ranked 1st: count/total.}
\label{tab:eval_combined}
\scriptsize
\renewcommand{\arraystretch}{1.1}
\resizebox{\columnwidth}{!}{%
\begin{tabular}{lccc|ccc|cccc}
\toprule
& \multicolumn{6}{c|}{\textbf{LLM-as-a-Judge}} & \multicolumn{4}{c}{\textbf{Human Experts}} \\
\cmidrule(lr){2-7} \cmidrule(lr){8-11}
\textbf{Method} & \multicolumn{3}{c|}{Avg.\ Score} & \multicolumn{3}{c|}{Ranked 1st} & \multicolumn{2}{c}{Avg.\ Score} & \multicolumn{2}{c}{Ranked 1st} \\
& Liver Cancer & eICU & Chinook & Liver Cancer & eICU & Chinook & Liver Cancer & eICU & Liver Cancer & eICU \\
\midrule
Direct Mapping~\cite{directmapsequeda2012directly} & 2.05 & 2.13 & 0.13 & 0/16  & 0/31  & 0/11  & 2.95 & 2.59 & 3/20  & 2/20 \\
Baseline~\cite{mateiu2023ontology}                 & 1.38 & 0.28 & 1.48 & 0/16  & 0/31  & 0/11  & 2.74 & 3.01 & 1/20  & 4/20 \\
Non-Iterative                                       & 1.89 & 0.67 & 2.77 & 0/16  & 0/31  & 0/11  & 2.39 & 1.24 & 2/20  & 0/20 \\
\textbf{RIGOR}                                      & \textbf{3.93} & \textbf{3.64} & \textbf{3.32} & \textbf{16/16} & \textbf{31/31} & \textbf{11/11} & \textbf{3.44} & \textbf{3.53} & \textbf{14/20} & \textbf{14/20} \\
\bottomrule
\end{tabular}
}
%\vspace
\scriptsize
\textit{Denominators for LLM-as-a-Judge Ranked 1st are 16 liver cancer tables, 31 eICU-CRD tables, and 11 Chinook tables. The human-expert denominator is 20 evaluations per medical dataset.}
\vspace{-2.5em}
\end{table}

\subsection{Ablation Study} \label{sec:ablation}

To isolate the contribution of each retrieval source, we ablate RIGOR on Chinook using Claude. The retrieval sources are: (1)~the relational schema and core-ontology context, (2)~external domain ontologies, and (3)~textual documentation. We evaluate five variants: no-retrieval, three single-source variants, and full RIGOR with all sources enabled. All variants retain the deterministic direct mapping; only the retrieved context varies. We evaluate alignment between the generated ontology and each retrieval resource using ontology classes/properties as candidates and resource elements as references: schema table/column names, external ontology class/property labels, and documentation sentences. Both sides are embedded with \texttt{all-MiniLM-L6-v2}; matches are counted when cosine similarity is at least 0.55, and we report precision, recall, and F1. Results in Table~\ref{tab:ablation} show that schema alignment is nearly saturated for all variants, since the direct-mapped backbone already preserves table and column names. Full RIGOR achieves the highest external-ontology recall and F1 and the highest average recall and F1 across all three resources, indicating the strongest cross-source alignment. Single-source variants provide targeted gains, e.g., documentation-only yields the highest documentation F1, but full RIGOR gives the best overall balance across schema, external ontologies, and documentation.

\section{Conclusions}

We presented \textbf{RIGOR}, an iterative RAG pipeline that
converts relational schemas into OWL2DL ontologies. A direct
mapping guarantees schema coverage; a \textit{Gen-LLM} enriches it using a growing core ontology, retrieved documentation, and external ontologies; a \textit{Judge-LLM} validates each delta ontology before integration. Experiments across three databases and two domains show that RIGOR consistently outperforms all competing methods, as confirmed by an \textit{LLM-as-a-Judge}, 10 human experts, and the BURR benchmark. 

\textbf{Limitations and future work.} Column coverage ranges from 0.77 to 1.00, indicating that some schema properties may be omitted during enrichment. Residual modeling pitfalls (e.g., \textit{P19}) reflect cases where the \emph{Judge-LLM} fails to catch \emph{Gen-LLM} errors; integrating tool-based feedback from validators such as OOPS! into the generation loop via agentic tool calling is a promising direction. RIGOR operates on schema and documentation by design, ensuring portability across deployments; however, when documentation is unavailable, instance-level signals (cardinality, value distributions, implicit key discovery) could help disambiguate cryptic columns. Coupling the \emph{Gen-LLM} with a symbolic reasoner may improve disjointness axioms~\cite{volker2015automatic}.

%We presented \textbf{RIGOR}, an iterative RAG pipeline that converts relational schemas into OWL2DL ontologies. A direct mapping guarantees schema coverage; a \textit{Gen-LLM} enriches it using growing core ontology, retrieved documentation, and external ontologies; a \textit{Judge-LLM} validates and refines each delta ontology before integration. Experiments across three databases spanning two domains show that RIGOR consistently outperforms other methods across all evaluation strategies, as confirmed by an \textit{LLM-as-a-Judge} and 10 human experts. \textbf{Limitations and future work.} RIGOR does not achieve perfect scores on all metrics. Column coverage ranges from 0.77 to 1.00 (Appendix \ref{eval_details}, Table \ref{tab:schema_coverage}), indicating that some schema properties may be omitted during semantic enrichment. Modeling pitfalls, while significantly reduced, are not eliminated entirely; residual pitfalls (e.g., multiple domains/ranges \textit{(P19)}) reflect cases where the \emph{Judge-LLM} fails to detect violations introduced by the \emph{Gen-LLM}. RIGOR operates on the schema and its documentation by design, ensuring portability across institutions sharing the same schema. However, incorporating instance-level data such as column cardinality, uniqueness, value distributions, and implicit primary/foreign-key discovery is a promising direction. Finally, disjointness axioms remain difficult to elicit from current LLMs; equipping the \emph{Gen-LLM} with a symbolic reasoner~\cite{volker2015automatic} may address this.

\section*{Acknowledgment}
The authors gratefully acknowledge support from the ATLAS project, funded by the German Federal Ministry of Education and Research (Bundesministerium für Bildung und Forschung, BMBF). The authors gratefully acknowledge the computing time provided on the HPC HoreKa by the National HPC Center at KIT (NHR@KIT). This center is jointly supported by the Federal Ministry of Education and Research and the Ministry of Science, Research, and the Arts of Baden-Württemberg as part of the National High-Performance Computing (NHR) joint funding program (https://www.nhr-verein.de/en/our-partners). HoreKa is partly funded by the German Research Foundation (DFG).

\section*{Supplemental Material Statement}
Source code for ontology generation methods, generated ontologies, RAG documents, SQL schemas, generated competency questions, and the online survey administered to experts are available at: \url{https://github.com/NadeenAhmad/RIGOR}.
%an anonymous repository: 

\section*{Declaration of use of Generative AI}
LLMs are used in our pipeline for ontology generation, fragment validation via a Judge-LLM, and post-hoc evaluation in competency question generation and LLM-as-Judge. For manuscript preparation, we used LLMs only for light editing, including rephrasing and improving the presentation of text, equations, and tables.
\bibliographystyle{splncs04}
\bibliography{ISWC26/main}

\section{Appendix}
This appendix provides supplementary material supporting the methods, evaluation, and implementation described in the main paper. It is organized as follows:

\begin{itemize}
    \item \textbf{RIGOR Formalization:} Formal definitions.
    %\item \textbf{External ontologies:} 
    \item \textbf{Experiments Details:} This subsection compares the capabilities of the considered ontology generation methods, showing differences between direct mapping\cite{directmapsequeda2012directly}, baseline\cite{mateiu2023ontology}, non-iterative generation methods, and our \textbf{RIGOR} pipeline. It also provides statistics on the external ontologies used in our experiments. The subsection also compares 
    \item \textbf{Prompt Templates:} Full prompts for ontology generation (RIGOR, baseline, non-iterative) and evaluation. 
    \item \textbf{Evaluation Details:} Examples of generated competency questions and answers, detailed results, and CQ-based LLM-as-a-Judge evaluation scores.
\end{itemize}

\subsection{RIGOR Formalization:}  \label{formalize}
\vspace{-2em}
\captionsetup{font=footnotesize}
\begin{table}[H]
\centering
\scriptsize
\renewcommand{\arraystretch}{1.2}
%\resizebox{0.65\columnwidth}{!}{%
\begin{tabular}{|c|l|}
\hline
\textbf{Symbol} & \textbf{Definition} \\ \hline
$\mathcal{S} = (\mathcal{R}, \key, \col, \FK)$ & Relational database schema \\ \hline
$\mathcal{R}$ & Set of relations (tables) in the database \\ \hline
$\key(r)$ & Primary-key attributes of relation $r$ \\ \hline
$\col(r)$ & Attributes of relation $r$ as (name, datatype) pairs \\ \hline
$\FK(r)$ & \makecell[l]{Foreign-key constraints of $r$: triples $(a, r', a')$ where attribute $a$ in $r$\\
references attribute $a'$ in $r'$} \\ \hline
$T = |\mathcal{R}|$ & Total number of iterations (one per table) \\ \hline
$\textsc{FK-BFS}(\mathcal{S})$ & BFS traversal of the FK dependency graph; returns table ordering $\langle r_1, \ldots, r_T \rangle$ \\ \hline
$\textit{docTable}(r)$ & Textual description of table $r$ \\ \hline
$\textit{docAttr}(r, a)$ & Textual description of attribute $a$ in table $r$ \\ \hline
$\mathcal{D}$ & Corpus of natural-language schema documentation \\ \hline
$\mathcal{E}$ & External ontology repository $\{\mathcal{O}_1^{E}, \dots, \mathcal{O}_n^{E}\}$ \\ \hline
$\mathcal{O}_t = (\mathcal{C}_t, \mathcal{P}_t, \mathcal{A}_t, \mathcal{M}_t)$ & Core ontology at iteration $t$ \\ \hline
$\mathcal{C}_t$ & Set of OWL class identifiers \\ \hline
$\mathcal{P}_t$ & Set of OWL property identifiers ($\mathcal{P}_t^{\mathrm{obj}} \cup \mathcal{P}_t^{\mathrm{data}}$) \\ \hline
$\mathcal{A}_t$ & Set of OWL2DL axioms \\ \hline
$\mathcal{M}_t$ & Set of annotation assertions (labels, comments, provenance) \\ \hline
$\text{clean}(r)$ & Preprocessing: removes ETL artefacts, corrects datatype mismatches \\ \hline
$\mathcal{DM}_r$ & Direct mapping for relation $r$ \\ \hline
$\Delta \mathcal{O}_r$ & Delta-ontology: semantic enrichment of $\mathcal{DM}_r$ \\ \hline
$\textsc{Validate}(\Delta\mathcal{O}_r, \mathcal{DM}_r, \mathcal{O}_{t-1})$ & Deterministic check: type consistency, domain/range cardinality, XSD ranges, schema coverage \\ \hline
$\textsc{Judge-LLM-Select}(\Delta\mathcal{O}_r, \mathcal{DM}_r)$ & Fallback: Judge-LLM selects best of delta ontology and direct mapping \\ \hline
$\Lex(\cdot)$ & Lexical view: human-readable strings (labels, comments) of an ontology or schema \\ \hline
$\mathcal{X}_t$ & Universe of texts available for embedding at iteration $t$ \\ \hline
$\mathcal{O}_{\mathrm{final}}$ & Final consolidated OWL ontology ($= \mathcal{O}_T$) \\ \hline
\end{tabular}%
%}
\caption{Formalization of relational database, textual descriptions, and ontology structures.}
\label{tab:formalization}
\end{table}

\subsection{Experiments Details:} \label{exp_details}

\begin{table}
\centering
\scriptsize
\renewcommand{\arraystretch}{1.25}

\resizebox{\columnwidth}{!}{%
\begin{tabular}{|l|c|c|c|c|}
\hline
\textbf{Capability} & 
\textbf{Direct Mapping \cite{directmapsequeda2012directly}} & 
\textbf{Baseline \cite{mateiu2023ontology}} & 
\textbf{Non-Iterative} & 
\textbf{RIGOR} \\
\hline
FK $\rightarrow$ ObjectProperty & 
\cmark\ deterministic & $\sim$ & $\sim$ & 
\cmark\ Direct mapping + Judge-LLM \\
\hline
\texttt{rdfs:label} / \texttt{rdfs:comment} & 
\xmark & \xmark & \cmark & \cmark \\
\hline
SubClassOf hierarchy & 
\xmark & \xmark & \xmark & \cmark \\
\hline
Alignment with external ontology& 
\xmark & \xmark & context only & \cmark \\
& 
 &  & (no alignment axioms) & \\
\hline
Value encoding annotations & 
\xmark & \xmark & \xmark & \cmark \\
\hline
%DisjointWith axioms & 
%\xmark & \xmark & \xmark & \cmark \\
%\hline
Existential restrictions & 
\xmark & \xmark & \xmark & \cmark \\
\hline
ETL artifact removal & 
\cmark & \xmark & \xmark & \cmark \\
\hline
Provenance annotations & 
\cmark & \xmark & \xmark & \cmark \\
\hline
Document retrieval & 
\xmark & \xmark & \cmark\ (FAISS) & \cmark  (FAISS) \\
\hline
Growing core ontology context & 
\xmark & \xmark & \xmark & \cmark \\
\hline
Complete column coverage & 
\cmark\ deterministic & $\sim$ & $\sim$ & 
\cmark\ Direct mapping \\
\hline
Judge-LLM validation & 
\xmark & \xmark & \xmark & \cmark \\
\hline
Iterative refinement & 
\xmark & \xmark & \xmark & \cmark \\
\hline
FK-guided traversal order & 
\xmark & \xmark & \xmark & \cmark \\
\hline
\end{tabular}%
}
\caption{Comparison of ontology generation methods by capability. 
\cmark\ = supported, \xmark\ = not supported, 
$\sim$ = prompted but LLM-dependent (no guarantee).}
\label{tab:method_comparison}
\end{table}

\begin{table}
\centering
\scriptsize
\setlength{\tabcolsep}{3pt}
\begin{tabular}{|l|r|r|r|r|}
\hline
\textbf{Name} & \textbf{\#Classes} & \textbf{\#Obj. Prop.} & \textbf{\#Data Prop.} & \textbf{\#Axioms} \\
\hline
Cell Ontology & 17,107 & 255 & 0 & 243,487 \\
Human Disease Ontology & 19,129 & 47 & 0 & 194,636 \\
ICD-10 & 12,445 & 233 & 1 & 17,139 \\
Ontology for Nutritional Studies & 4,735 & 73 & 3 & 27,326 \\
\hline     
Music Ontology \footref{music_onto} & 12,784 & 499 & 14 & 66,088 \\
Performed Music Ontology \footref{performed_music} & 45 & 28 & 9 & 313 \\
DOREMUS Ontology \footref{dorem_onto} & 211 & 656 & 59 & 7,583 \\
\hline
\end{tabular}
\caption{Statistics of selected external ontologies, including biomedical ontologies from BioPortal\footref{bioportal} and music-domain ontologies.}
\label{struct}
\end{table}

\begin{table}
\centering
\scriptsize
\renewcommand{\arraystretch}{1.25}
%\resizebox{\columnwidth}{!}{%
\begin{tabular}{|l|c|c|c|}
\hline
\textbf{Variant} & \textbf{Schema Context} & \textbf{External Ontologies} & \textbf{Relevant Documents} \\
\hline
no\_rag              & \xmark        & \xmark       & \xmark       \\
\hline
only\_schema         & \cmark     & \xmark        & \xmark        \\
\hline
only\_external       & \xmark        & \cmark     & \xmark        \\
\hline
only\_docs           & \xmark        & \xmark        & \cmark     \\
\hline

full\_RIGOR           & \cmark     & \cmark     & \cmark     \\
\hline
\end{tabular}%
%}
\caption{Ablation study configurations. All variants retain the deterministic direct mapping to maintain schema coverage. \cmark\ = enabled, \xmark = disabled.}
\label{tab:ablation-variants}
\end{table}

\subsection{Prompt Templates:} \label{box:prompt-example}
This section presents the prompts used for ontology generation and evaluation. \textcolor{blue}{Blue text} denotes dynamic input that is populated at runtime from the pipeline (e.g., relational schemas, retrieved context, direct mappings); all remaining text is a static instruction. For ontology generation, we include the prompts used in \textbf{RIGOR}—namely \hyperref[promptmain]{(A.1)} the \emph{Gen-LLM} prompt and \hyperref[fig:judgeLLM]{(A.2)} the \emph{Judge-LLM} prompt—as well as \hyperref[fig:baseline]{(A.3)} the prompt used for the baseline approach and \hyperref[fig:noniterative_prompt]{(A.4)} the prompt used for the non-iterative generation method.
For evaluation, we provide \hyperref[fig:cq_gen]{(B.1)} the prompt for competency question generation and \hyperref[fig:llm-As-Judge]{(B.2)} the LLM-as-a-Judge prompt used to assess ontology fragments. \\

\begin{tcolorbox}[
  colback=gray!10,
  colframe=black,
  boxrule=0.6pt,
  arc=6pt,
  left=4pt, right=4pt, top=6pt, bottom=6pt,
  breakable,
  title=A.1: RIGOR \textit{Gen-LLM} Prompt,
  colbacktitle=green!20,
  coltitle=black,
  fonttitle=\bfseries
]
  \label{promptmain}
\scriptsize

You are receiving a direct mapping draft for database table \texttt{'\{table\_name\}'}. Your role is semantic enrichment; do not redeclare what is already there unless you are upgrading it, and do not change existing property or class names.

\vspace{0.5em}
\textbf{[CONTEXT]} \vspace{0.5em}\\
- Direct Mapping Draft for this table: \\
\textcolor{blue}{\texttt{'\{direct\_mapping\}'}}\\
- Database Schema (current table + FK targets): \\
\textcolor{blue}{\texttt{'\{schema\_str\}'}}\\
- Foreign Key Constraints for \textcolor{blue}{\texttt{'\{table\_name\}'}}: \\
\textcolor{blue}{\texttt{'\{foreign\_keys\_str\}'}}\\
- Relevant Documents: \\
\textcolor{blue}{\texttt{'\{documents\}'}}\\
- Existing Core Ontology Knowledge (already built classes/properties): \\
\textcolor{blue}{\texttt{'\{existing\_ontology\}'}}\\

\vspace{0.5em}
\textbf{[MANDATORY TASKS]}\vspace{0.5em}\\
You MUST perform ALL of the following. These are not optional.\\

\textbf{1. LABELS AND COMMENTS (Required):}\\
Add \texttt{rdfs:label} in English and \texttt{rdfs:comment} to every class and property.\\
For abbreviated or non-English names, expand them in the label and explain them in the comment. \\
For properties encoding units in their name (e.g., \texttt{radiation\_amount\_in\_gray}, \texttt{ASAT\_\textmu mol\_per\_l}), document the unit in the comment.\\

\textbf{2. FOREIGN KEY RESOLUTION (Required):}\\
For every column listed under Foreign Key Constraints above, replace its DataProperty with an \texttt{owl:ObjectProperty} pointing to the referenced class.\\
Remove the original DataProperty; do not keep both.\\
Example: \texttt{patient\_id} as a FK to \texttt{PatientData} becomes a \texttt{hasPatientData} ObjectProperty; the \texttt{patient\_id} DataProperty must then be dropped entirely.\\

\textbf{3. SUBCLASSOF HIERARCHY (Required):}\\
Add \texttt{SubClassOf} axioms where a natural specialization exists.\\
Use only fully qualified OWL IRIs — \texttt{owl:Thing} must be referenced as \texttt{http://www.w3.org/2002/07/owl\#Thing}, never as a local namespace alias.\\

\textbf{4. ANNOTATIONS FOR ENCODED VALUES (Required):}\\
If a DataProperty has \texttt{xsd:integer} range but represents a categorical or ordinal concept (e.g., \texttt{ecog\_stage}, \texttt{asa\_score}, \texttt{her\_2\_status}), add an \texttt{rdfs:comment} documenting the value encoding.\\

\textbf{5. DISJOINTNESS (Required):}\\
Declare \texttt{DisjointWith} axioms between sibling classes that cannot share instances.\\

\textbf{6. CARDINALITY CONSTRAINTS (Required):}\\
For any column declared as not-null, unique, or a primary key, assert an existential restriction (\texttt{C SubClassOf ($\exists$ property value)}) on its property.\\

\vspace{0.5em}
\textbf{[STRICT CONSTRAINTS]}\vspace{0.5em}\\
\textbf{TYPE CONSISTENCY:} A URI must not be declared as both \texttt{owl:ObjectProperty} and \texttt{owl:DatatypeProperty}.\\
\textbf{DOMAIN AND RANGE:} Every property MUST declare exactly one \texttt{rdfs:domain} and one \texttt{rdfs:range}.\\
\textbf{STANDARD DATATYPES:} All DataProperty ranges must use XSD datatypes.\\
\textbf{NO REDUNDANT FK DATAPROPERTIES:} Once a FK column is resolved to an ObjectProperty, the original integer DataProperty for that column must not appear in the output.\\
\textbf{CORRECT OWL IRIS:} \\ \texttt{SubClassOf owl:Thing} must use \texttt{http://www.w3.org/2002/07/owl\#Thing}.\\
\textbf{NO SELF-REFERENTIAL PROPERTIES:} Do not declare an ObjectProperty whose domain and range are the same class unless there is an explicit, meaningful recursive relationship.\\

\vspace{0.5em}
\textbf{[EXAMPLE FORMAT]}\vspace{0.5em}\\
\textbf{Class:} \texttt{'\{table\_name\}'}\\
Annotations: \\
{\scriptsize\texttt{
rdfs:label "Human-readable class name"@en,\\
rdfs:comment "Description of what this class represents clinically"@en,\\
prov:wasDerivedFrom <http://example.org/provenance/'\{table\_name\}'>
}}\\
\vspace{0.3em}

\textbf{DataProperty:} \texttt{has\_column\_name}\\
\texttt{Domain '\{table\_name\}'}\\
\texttt{Range xsd:datatype}\\
Annotations: \\
{\scriptsize\texttt{
rdfs:label "Human-readable property name"@en,\\
rdfs:comment "Description including units or encoding if applicable"@en,\\
prov:wasDerivedFrom <http://example.org/provenance/'\{table\_name\}'/column\_name>
}}\\
\vspace{0.3em}

\textbf{ObjectProperty:} \texttt{relatedClass}\\
\texttt{Domain '\{table\_name\}'}\\
\texttt{Range ReferencedClass}\\
Annotations: \\
{\scriptsize\texttt{
rdfs:label "Human-readable property name"@en,\\
rdfs:comment "Description of the relationship"@en,\\
prov:wasDerivedFrom <http://example.org/provenance/'\{table\_name\}'/fk\_column>
}}\\

\vspace{0.3em}
Output ONLY Manchester Syntax and nothing else. \\ \textbf{[OUTPUT]}
\end{tcolorbox}
\vspace{-1.5em}
\captionof{figure}{A.1 -- \textbf{RIGOR} \emph{Gen-LLM} prompt: The prompt provides the model with a direct mapping, foreign key constraints, textual documentation, and the current state of the \emph{core ontology}. It instructs the model to perform semantic enrichment by adding labels and comments, resolving foreign keys into object properties, removing non-semantic artifacts, and introducing additional axioms such as subclass relations, disjointness, and cardinality constraints, while adhering to strict modeling requirements (e.g., type consistency).}
\label{fig:promptbox}

\vspace{2em}

\begin{tcolorbox}[
  colback=gray!10,
  colframe=black,
  boxrule=0.6pt,
  arc=6pt,
  left=4pt, right=4pt, top=6pt, bottom=6pt,
  breakable,
  title=A.2: RIGOR \emph{Judge-LLM} Prompt,
  colbacktitle=green!20,
  coltitle=black,
  fonttitle=\bfseries
]
\scriptsize
You are an ontology validation expert. Evaluate the following delta ontology fragment generated for database table \textcolor{blue}{\texttt{'\{table\_name\}'}}. Your task is to assess it against the criteria below and either approve it or return it with specific correction instructions. \\

%\textbf{[INPUT]}\\
- Direct Mapping Draft for this table:
\textcolor{blue}{\texttt{'\{direct\_mapping\}'}}

- Delta Ontology Fragment to validate:
\textcolor{blue}{\texttt{'\{delta\_ontology\}'}}

- Database Schema for this table:
\textcolor{blue}{\texttt{'\{schema\_context\}'}}

- Foreign Key Constraints:
\textcolor{blue}{\texttt{'\{foreign\_keys\_str\}'}}

- Current Core Ontology (relevant excerpt):
\textcolor{blue}{\texttt{'\{existing\_ontology\}'}} \\

\vspace{0.5em}
\textbf{[VALIDATION CRITERIA]}\\

\textbf{1. FOREIGN KEY COMPLETENESS (Critical):}\\
   For every column listed in the foreign key constraints, verify that an \texttt{owl:ObjectProperty} exists in the delta ontology with the correct domain (this class)
   and range (the referenced class). \\
   If any FK column still appears as a DataProperty with \texttt{xsd:integer} range, reject the fragment and instruct the Gen-LLM to resolve it.\\
\vspace{0.5em}

\textbf{2. NO REDUNDANT FK DATAPROPERTIES (Critical):}\\
   Verify that no DataProperty exists for a column that has already been resolved to an ObjectProperty. If both exist for the same FK column, instruct the removal of the DataProperty. \\

\textbf{3. NO SELF-REFERENTIAL OBJECTPROPERTIES (Critical):}\\
   Verify that no ObjectProperty has the same class as both its domain and its range, unless a meaningful recursive relationship is explicitly documented. \\

\textbf{4. DIRECT MAPPING COVERAGE (Critical):}\\
   Compare the delta ontology against the direct mapping draft. Verify that every column present in the direct mapping draft is either represented as a DataProperty, resolved to an ObjectProperty, or explicitly justified for exclusion (e.g., ETL artifact or derived property). Any column silently dropped without justification must be flagged, and the fragment rejected. \\

\textbf{5. CARDINALITY CONSTRAINTS (Critical):}\\
   Verify that every column declared as not-null, unique, or a primary key in the source schema has a corresponding existential restriction or qualified cardinality restriction. If missing, instruct the Gen-LLM to add them.\\

\textbf{6. DISJOINTNESS (Important):}\\
   Verify that sibling classes sharing a common superclass carry DisjointWith axioms where instances cannot belong to more than one sibling class simultaneously. \\

\textbf{7. ENCODED CATEGORICALS (Important):}\\
   Check whether any DataProperty with \texttt{xsd:integer} range represents a categorical or
   ordinal concept. If so, verify that either an enumerated class models it or an \texttt{rdfs:comment} documents the encoding. \\

\textbf{8. ANNOTATION COMPLETENESS (Important):}\\
   Verify that every class and property carries \texttt{rdfs:label} in English and \texttt{rdfs:comment}. Flag any element with a non-English, abbreviated, or cryptic local name that lacks
   an explanatory label. Flag any property whose name encodes a unit of measurement without a corresponding unit annotation or \texttt{rdfs:comment}. \\

\textbf{9. CORE ONTOLOGY CONSISTENCY (Important):}\\
   Verify that the delta ontology does not redefine or conflict with classes or properties already present in the core ontology. If a concept already exists, the delta should reuse or subclass it rather than introduce a duplicate.\\

\textbf{10. JUNCTION TABLE HANDLING (Important):}\\
    If the table being processed is a pure junction table (only two FK columns and no independent attributes), verify that it has been modeled as an ObjectProperty
    between the two referenced classes rather than as a standalone class.\\

\textbf{11. PROPERTY TYPE CORRECTNESS (Important):}\\
    Verify that no URI is declared as both \texttt{owl:ObjectProperty} and \texttt{owl:DatatypeProperty}.
    Verify that all DataProperty ranges use valid XSD datatypes. Verify that date and time properties use \texttt{xsd:date} or \texttt{xsd:dateTime}, not \texttt{xsd:float} or \texttt{xsd:string}.\\

\textbf{12. EXTERNAL ALIGNMENT (Minor):}\\
    Verify that at least one class or property in the delta carries an external alignment annotation (\texttt{owl:equivalentClass}, \texttt{rdfs:subClassOf} to an external IRI, or \texttt{skos:exactMatch}) where a corresponding concept plausibly exists in a standard biomedical ontology.\\

\textbf{13. HIERARCHY (Minor):}\\
    Verify that \texttt{SubClassOf} axioms are present where a natural specialization relationship exists between this class and another class in the core ontology. \\

\textbf{14. PROVENANCE (Minor):}\\
    Verify that every class and property carries a \texttt{prov:wasDerivedFrom} annotation referencing its source table or column. \\

\textbf{[OUTPUT FORMAT]} \\

Return your assessment using EXACTLY this structure (no extra text outside it):

Decision: APPROVED | REJECTED | APPROVED\_WITH\_CORRECTIONS \\
- Critical issues (must fix before merging): [list each issue on a new line, or "none"] \\
- Important issues (should fix before merging): [list each issue on a new line, or "none"] \\
- Minor issues (can fix in a later pass): [list each issue on a new line, or "none"]
\vspace{0.5em}\\

Corrected fragment: [full corrected OWL Manchester Syntax if Decision is not APPROVED, otherwise "N/A"]

\end{tcolorbox}
\vspace{-1.5em}
\captionof{figure}{A.2 -- \emph{Judge-LLM} prompt for validation and refinement of ontology fragments in \textbf{RIGOR}. The model receives as input the direct mapping draft, the generated \emph{delta ontology}, the database schema, the foreign key constraints, and the current \emph{core ontology}, and evaluates the fragment against structural and semantic criteria (e.g., foreign key resolution, coverage, cardinality constraints, consistency, and annotation completeness). It outputs a structured decision (APPROVED, REJECTED, or APPROVED\_WITH\_CORRECTIONS), categorized issues (critical, important, minor), and, if necessary, a corrected fragment in Manchester Syntax.}
\label{fig:judgeLLM}

\vspace{2em}

\begin{tcolorbox}[
  colback=gray!10,
  colframe=black,
  boxrule=0.6pt,
  arc=6pt,
  left=4pt, right=4pt, top=6pt, bottom=6pt,
  title=A.3: Baseline Generation Prompt \cite{mateiu2023ontology},
  colbacktitle=green!20,
  coltitle=black,
  fonttitle=\bfseries
]\scriptsize
Generate an OWL 2 ontology fragment in Manchester Syntax for the database table 
\textcolor{blue}{\texttt{'\{table\_name\}'}}, Table schema: \textcolor{blue}{\texttt{'\{schema\_str\}'}}\\

Instructions:\\
- Create an OWL Class for the table\\
- Create a DataProperty for each non-FK column with correct domain and range\\
- Create an ObjectProperty for each FK column\\
- Use xsd: datatypes (xsd:string, xsd:integer, xsd:boolean, xsd:float, xsd:dateTime)\\
- Every property MUST have exactly one Domain and one Range\\
- Output ONLY valid Manchester Syntax, nothing else\\

\textbf{[OUTPUT]}
\end{tcolorbox}
\vspace{-1.5em}
\captionof{figure}{A.3 -- Baseline generation prompt \cite{mateiu2023ontology}.}
\label{fig:baseline}

 \begin{tcolorbox}[
  colback=gray!10,
  colframe=black,
  boxrule=0.6pt,
  arc=6pt,
  left=4pt, right=4pt, top=6pt, bottom=6pt,
  breakable,
  title=A.4: Non-Iterative Generation Prompt,
  colbacktitle=green!20,
  coltitle=black,
  fonttitle=\bfseries
]
\scriptsize

Generate an OWL~2 ontology fragment in Manchester Syntax for the database table \textcolor{blue}{\texttt{'\{table\_name\}'}}.\\

\vspace{0.5em}
\textbf{[SCHEMA OF CURRENT TABLE]}\\
\textcolor{blue}{\texttt{'\{col\_lines\}'}}\\

\vspace{0.5em}
\textbf{[FOREIGN KEYS]}\\
\textcolor{blue}{\texttt{'\{fk\_lines\}'}}\\

\vspace{0.5em}
\textbf{[FULL DATABASE SCHEMA CONTEXT]}\\
\textcolor{blue}{\texttt{'\{schema\_str\}'}}\\

\vspace{0.5em}
\textbf{[RELEVANT DOCUMENTATION]}\\
\textcolor{blue}{\texttt{'\{docs\_text\}'}}\\

\vspace{0.5em}
\textbf{[RELEVANT EXTERNAL ONTOLOGY CONCEPTS]}\\
\textcolor{blue}{\texttt{'\{onto\_text\}'}}\\  

\vspace{0.5em}
\textbf{[INSTRUCTIONS]}\\
- Create an OWL Class for this table\\
- For every FK column: create an \texttt{owl:ObjectProperty} linking to the referenced class\\
- For every non-FK column: create a \texttt{owl:DatatypeProperty} with correct \texttt{xsd:} range\\
- Add \texttt{rdfs:label} and \texttt{rdfs:comment} to every class and property\\
- Every property MUST have exactly one Domain and one Range\\
- Output ONLY valid Manchester Syntax, nothing else\\

\vspace{0.5em}
\textbf{[OUTPUT]}

\end{tcolorbox}
%\vspace{-3.5em}

\captionof{figure}{A.4 -- Non-iterative generation prompt: The model receives the current table schema together with its foreign keys, the full database schema as global context, relevant documentation, and retrieved external ontology concepts. It generates a table-specific ontology fragment directly, without iterative refinement or Judge-LLM validation, and outputs only Manchester Syntax.}
\label{fig:noniterative_prompt}

\vspace{2em}

\begin{tcolorbox}[
  colback=gray!10,
  colframe=black,
  boxrule=0.6pt,
  arc=4pt,
  left=3pt, right=3pt, top=3pt, bottom=3pt,
  title=\textbf{B.1: CQ Generation Prompt},
  colbacktitle=blue!20,
  coltitle=black,
  fonttitle=\bfseries
]
\scriptsize
Given the SQL table schema for table \textcolor{blue}{\texttt{\{table\_name\}}}:\\
Generate 5 competency questions (CQs) that this table's ontology should answer.\\
For each question, also provide a short answer explaining how the ontology
would answer it using the table schema. Think step-by-step.\\[2pt]
Competency Questions for Table \textcolor{blue}{\texttt{\{table\_name\}}}:\\
\textbf{1. [Question]}\\
\textbf{Answer: [Explanation]}\\
Continue similarly for 5 questions. Output only this format.
\end{tcolorbox}

\captionof{figure}{B.1 -- CQ generation prompt: produces 5 table-specific CQs and short ontology-based answers using fixed delimiters (e.g., \texttt{[Question]}, \texttt{Answer:}) and the runtime table name \texttt{\{table\_name\}}.}
\label{fig:cq_gen}

\begin{tcolorbox}[
  colback=gray!10,
  colframe=black,
  boxrule=0.6pt,
  arc=4pt,
  left=3pt, right=3pt, top=3pt, bottom=3pt,
  title=\textbf{B.2: LLM-as-a-Judge Evaluation Prompt},
  colbacktitle=blue!20,
  coltitle=black,
  fonttitle=\bfseries
]
\scriptsize
You are an ontology evaluation expert. Below are four OWL ontology fragments (A, B, C, D) generated for the same database table by different methods, presented anonymously. You also receive the source table schema and 5 competency questions (CQs).\\[2pt]
\textbf{Task:}\\
1. RANK ontologies A--D from best (1st) to worst (4th).\\
2. SCORE each ontology on a 0--5 scale for each of the seven dimensions defined below.\\
3. Provide a BRIEF justification (1 sentence) for the top-ranked and bottom-ranked ontology.\\[2pt]
Table Schema: \textcolor{blue}{\texttt{\{schema\}}}\\
Competency Questions: \textcolor{blue}{\texttt{\{cqs\}}}\\[2pt]
Ontology A: \textcolor{blue}{\texttt{\{onto\_a\}}}\\
Ontology B: \textcolor{blue}{\texttt{\{onto\_b\}}}\\
Ontology C: \textcolor{blue}{\texttt{\{onto\_c\}}}\\
Ontology D: \textcolor{blue}{\texttt{\{onto\_d\}}}\\[2pt]
\textbf{Output:} Respond with ONLY valid JSON containing: table name, ranking (array of letters), scores (per letter, per dimension), best/worst letter with a sentence reason.
\end{tcolorbox}
\vspace{-1.2em}
\captionof{figure}{B.2 -- LLM-as-a-Judge prompt: four anonymized ontology fragments (A = W3C Direct Mapping, B = Baseline, C = Non-Iterative, D = RIGOR) are scored side-by-side across seven quality dimensions, with the table schema and CQs as grounding context. Method identities are withheld from the judge during evaluation.}
\label{fig:llm-As-Judge}

\vspace{1em}
\noindent\textbf{Evaluation Dimensions.}
The \textit{LLM-as-a-Judge} and human experts evaluate ontology fragments along seven dimensions, grounded in established ontology quality frameworks~\cite{article,vrandevcic2009ontology,tartir2005ontoqa}:
\begin{itemize}
    \item \textbf{Accuracy:} Correct representation of the domain; classes and properties faithfully reflect the table schema and its semantics.
    \item \textbf{Completeness:} All significant columns and relationships are represented, with annotations (\texttt{rdfs:label}, \texttt{rdfs:comment}) on every element. Value encodings are documented for coded integer columns.
    \item \textbf{Conciseness:} No redundant or unnecessary elements; every class and property serves a purpose.
    \item \textbf{Adaptability:} \texttt{SubClassOf} hierarchy enables extension; alignment links to established external vocabularies exist.
    \item \textbf{Clarity:} Every element has a human-readable label and comment. Abbreviated names are expanded; units are documented where applicable
    \item \textbf{Consistency:} No logical contradictions; no punning (same URI as both object and data property); correct property types; \texttt{rdfs:domain} and \texttt{rdfs:range} present on every property; FK columns resolved to \texttt{owl:ObjectProperty}, not left as \texttt{owl:DatatypeProperty}.
    \item \textbf{Domain Enrichment:} The degree to which the ontology transcends structural schema mirroring to capture domain-level semantics: (a)~annotations that explain the clinical or domain meaning of classes and properties, including expansion of abbreviated names and documentation of measurement units; (b)~value encoding annotations that explain the meaning of coded integer values (e.g., ECOG 0--5 scale); (c)~class hierarchies (\texttt{SubClassOf}) reflecting established domain taxonomies; (d)~provenance metadata linking each element to its source table or documentation. %~\cite{vrandevcic2009ontology,tartir2005ontoqa}
\end{itemize}

\subsection{Evaluation Details:}\label{eval_details}
\begin{figure*}
    \centering
    \begin{subfigure}[t]{0.49\textwidth}
        \centering
        \includegraphics[width=\linewidth]{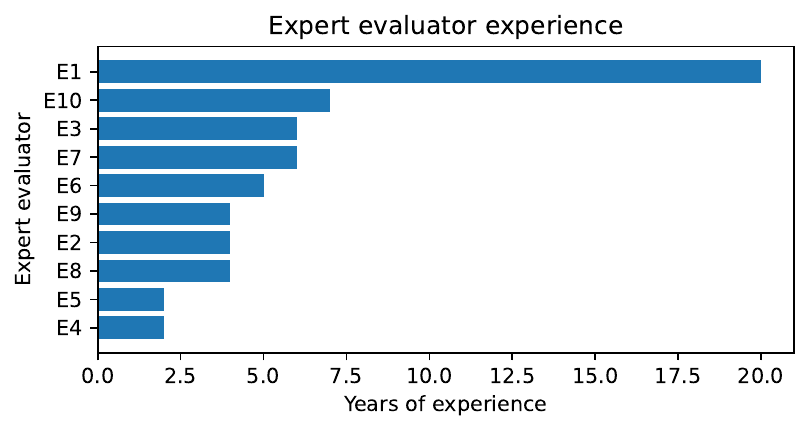}
        \caption{}
        \label{fig:expert_years}
    \end{subfigure}
    \hfill
    \begin{subfigure}[t]{0.49\textwidth}
        \centering
        \includegraphics[width=\linewidth]{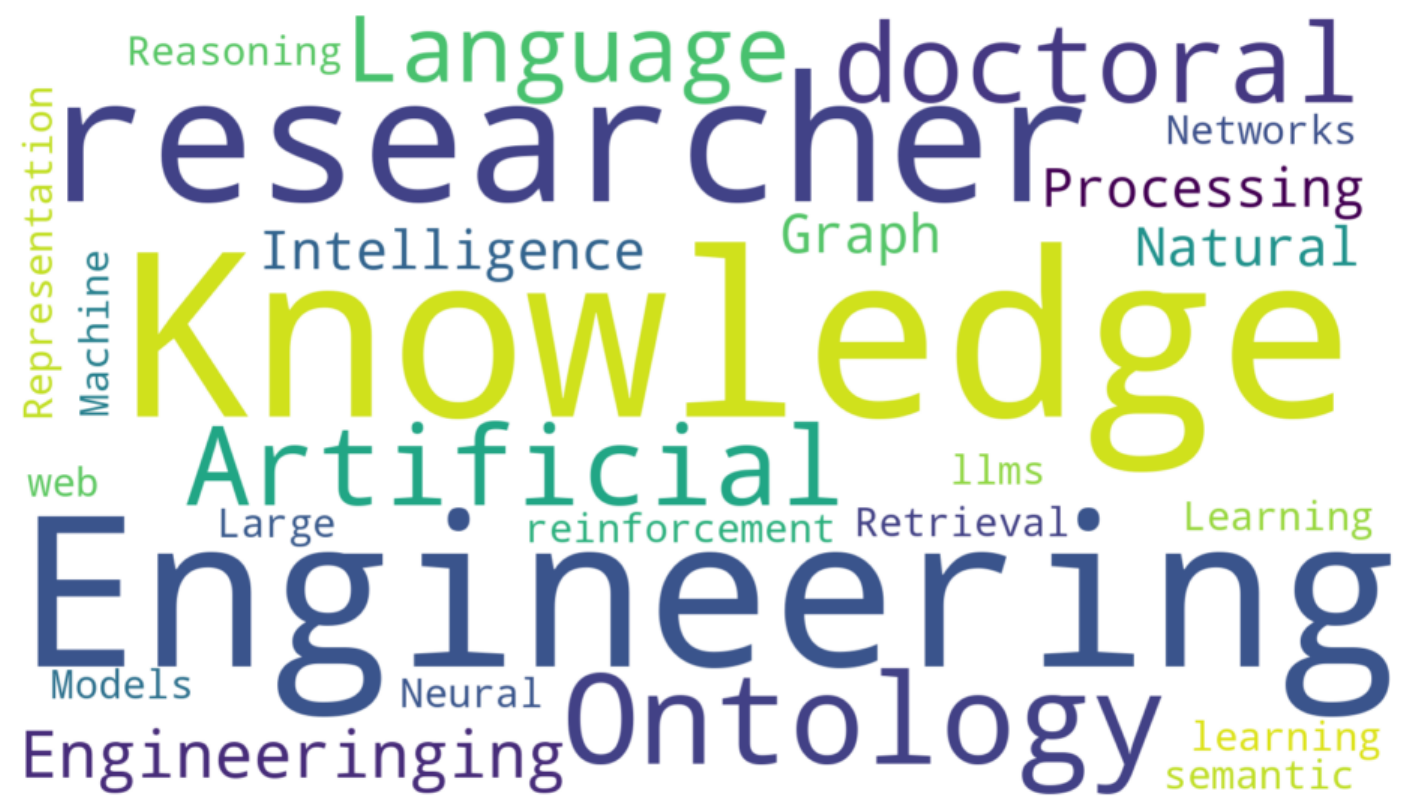}
        \caption{}
        \label{fig:expert_wordcloud}
    \end{subfigure}
    \caption{Profile of the expert evaluators participating in the survey. (a): years of experience for each expert (E1--E10). (b): word cloud summarizing the evaluators' self-reported areas of expertise.}
    \label{fig:expert_profile}
\end{figure*}

\begin{table}
\centering
\caption{Examples of generated competency questions for the \texttt{physicalexam} table from the eICU-CRD \cite{article}.}
\label{tab:physicalexam-cqs}
\scriptsize
\begin{tabular}{p{0.49\linewidth} p{0.49\linewidth}}
\toprule
\textbf{Competency Question} & \textbf{Expected Answer} \\
\midrule
What physical examination findings were documented for a given patient unit stay?
&
The physical examination findings associated with the specified patient unit stay, including the examined body system and the corresponding clinical finding.
\\
\addlinespace
Which patients had abnormal physical examination findings documented in a specific clinical domain, such as respiratory, neurological, or cardiovascular assessment?
&
The patient unit remains with the physical examination records in the specified clinical domain, with findings indicating abnormal status.
\\
\addlinespace
How do physical examination findings relate to the patient's vital signs and laboratory results at a comparable time point during the ICU stay?
&
The physical examination findings, linked to vital signs and laboratory results for the same patient unit stay, are comparable at comparable observation times, supporting an integrated view of the patient's clinical state.
\\
\bottomrule
\end{tabular}
\end{table}

\begin{sidewaystable}
\centering
\caption{OOPS! pitfall counts. ``*'' denotes a global pitfall reported without an exact count. (Cl = Claude, Mi = Mistral, DS = DeepSeek). OOPS! scans for 41 pitfalls, categorized as \textit{critical}, \textit{important}, or \textit{minor} based on their impact on ontology quality.}
\label{tab:pitfalls}
\scriptsize
\setlength{\tabcolsep}{2.4pt}
\renewcommand{\arraystretch}{1.02}
\resizebox{\columnwidth}{!}{%
\begin{tabular}{lc|cccccccccc|cccccccccc|cccccccccc}
\toprule
& & \multicolumn{10}{c|}{\textbf{Liver Cancer}} & \multicolumn{10}{c|}{\textbf{eICU-CRD}} & \multicolumn{10}{c}{\textbf{Chinook}} \\
\cmidrule(lr){3-12}\cmidrule(lr){13-22}\cmidrule(l){23-32}
\textbf{Pitfall} & \textbf{Sev.}
& \textbf{\shortstack{Direct\\Mapping \cite{directmapsequeda2012directly}}} & \multicolumn{3}{c}{\textbf{Baseline \cite{mateiu2023ontology}}} & \multicolumn{3}{c}{\textbf{\shortstack{Non-\\Iterative}}} & \multicolumn{3}{c|}{\textbf{RIGOR}}
& \textbf{\shortstack{Direct\\Mapping \cite{directmapsequeda2012directly}}} & \multicolumn{3}{c}{\textbf{Baseline \cite{mateiu2023ontology}}} & \multicolumn{3}{c}{\textbf{\shortstack{Non-\\Iterative}}} & \multicolumn{3}{c|}{\textbf{RIGOR}}
& \textbf{\shortstack{Direct\\Mapping \cite{directmapsequeda2012directly}}} & \multicolumn{3}{c}{\textbf{Baseline \cite{mateiu2023ontology}}} & \multicolumn{3}{c}{\textbf{\shortstack{Non-\\Iterative}}} & \multicolumn{3}{c}{\textbf{RIGOR}} \\
& 
& & Cl & Mi & DS & Cl & Mi & DS & Cl & Mi & DS
& & Cl & Mi & DS & Cl & Mi & DS & Cl & Mi & DS
& & Cl & Mi & DS & Cl & Mi & DS & Cl & Mi & DS \\
\midrule
\textbf{Total} & --
& 731 & 422 & 397 & 403 & 29 & 61 & 18 & \textbf{20} & \textbf{23} & \textbf{3}
& 665 & 434 & 440 & 421 & 39 & 220 & 27 & \textbf{2} & \textbf{2} & \textbf{26}
& 102 & 95 & 74 & 81 & 24 & 35 & 14 & \textbf{0} & \textbf{0} & \textbf{0} \\
\midrule
P04 & M
& 0   & 0   & 0   & 0   & 5   & 2   & 3   & \textbf{3}   & \textbf{3}   & \textbf{0}
& 0   & 0   & 0   & 0   & 8   & 1   & 0   & \textbf{0}   & \textbf{0}   & \textbf{2}
& 0   & 0   & 0   & 0   & 0   & 0   & 0   & \textbf{0} & \textbf{0} & \textbf{0} \\
P08 & M
& 368 & 400 & 379 & 387 & 0   & 30  & 0   & \textbf{3}   & \textbf{0}   & \textbf{0}
& 332 & 385 & 364 & 382 & 0   & 154 & 2   & \textbf{0}   & \textbf{0}   & \textbf{2}
& 47  & 60  & 49  & 55  & 0   & 19  & 0   & \textbf{0} & \textbf{0} & \textbf{0} \\
P10 & I
& *   & *   & *   & *   & *   & *   & *   & \textbf{*}   & \textbf{*}   & \textbf{*}
& *   & *   & *   & *   & *   & *   & *   & \textbf{*}   & \textbf{*}   & \textbf{*}
& *   & *   & *   & *   & *   & *   & *   & \textbf{0} & \textbf{0} & \textbf{0} \\
P11 & I
& 345 & 0   & 0   & 0   & 0   & 0   & 0   & \textbf{0}   & \textbf{0}   & \textbf{0}
& 299 & 0   & 0   & 0   & 0   & 0   & 0   & \textbf{0}   & \textbf{0}   & \textbf{0}
& 27  & 0   & 0   & 0   & 0   & 0   & 0   & \textbf{0} & \textbf{0} & \textbf{0} \\
P12 & I
& 0   & 5   & 2   & 2   & 2   & 0   & 0   & \textbf{0}   & \textbf{0}   & \textbf{0}
& 2   & 6   & 2   & 5   & 1   & 1   & 3   & \textbf{0}   & \textbf{0}   & \textbf{2}
& 0   & 1   & 0   & 0   & 0   & 0   & 0   & \textbf{0} & \textbf{0} & \textbf{0} \\
P13 & M
& 7   & 6   & 2   & 3   & 10  & 16  & 7   & \textbf{7}   & \textbf{7}   & \textbf{0}
& 2   & 18  & 23  & 8   & 24  & 25  & 7   & \textbf{0}   & \textbf{0}   & \textbf{2}
& 9   & 11  & 6   & 7   & 20  & 13  & 9   & \textbf{0} & \textbf{0} & \textbf{0} \\
P19 & C
& 6   & 9   & 10  & 8   & 2   & 3   & 1   & \textbf{5}   & \textbf{7}   & \textbf{0}
& 25  & 23  & 32  & 20  & 2   & 27  & 10  & \textbf{0}   & \textbf{0}   & \textbf{13}
& 14  & 20  & 15  & 17  & 0   & 1   & 3   & \textbf{0} & \textbf{0} & \textbf{0} \\
P20 & M
& 0   & 0   & 0   & 0   & 0   & 0   & 0   & \textbf{0}   & \textbf{4}   & \textbf{0}
& 0   & 0   & 0   & 0   & 0   & 0   & 0   & \textbf{0}   & \textbf{0}   & \textbf{0}
& 0   & 0   & 0   & 0   & 0   & 0   & 0   & \textbf{0} & \textbf{0} & \textbf{0} \\
P22 & M
& *   & *   & *   & *   & *   & *   & *   & \textbf{*}   & \textbf{*}   & \textbf{*}
& *   & *   & *   & *   & *   & *   & *   & \textbf{*}   & \textbf{*}   & \textbf{*}
& *   & *   & *   & *   & *   & *   & *   & \textbf{0} & \textbf{0} & \textbf{0} \\
P32 & M
& 0   & 0   & 0   & 0   & 8   & 0   & 2   & \textbf{0}   & \textbf{0}   & \textbf{1}
& 0   & 0   & 0   & 0   & 2   & 0   & 0   & \textbf{0}   & \textbf{0}   & \textbf{0}
& 0   & 0   & 0   & 0   & 2   & 0   & 0   & \textbf{0} & \textbf{0} & \textbf{0} \\
P34 & I
& 3   & 0   & 2   & 1   & 0   & 8   & 3   & \textbf{0}   & \textbf{0}   & \textbf{0}
& 3   & 0   & 17  & 4   & 0   & 10  & 3   & \textbf{0}   & \textbf{0}   & \textbf{3}
& 3   & 1   & 2   & 0   & 0   & 0   & 0   & \textbf{0} & \textbf{0} & \textbf{0} \\
\bottomrule
\end{tabular}
}
\vspace{0.3em}
\raggedright
\scriptsize \textit{Sev. = Severity: C = critical, I = important, M = minor. Pitfalls: P04 unconnected classes; P08 missing annotations; P10 missing disjointness; P11 missing domain/range; P12 equivalent properties not declared; P13 inverse properties not declared; P19 multiple domains/ranges; P20 misusing ontology annotations; P22 naming inconsistency; P32 missing inverse in relation; P34 untyped class.}
\end{sidewaystable}

\begin{table}
\centering
\caption{Structural ontology metrics across methods and models. LogicalAx = logical axioms, AnnAx = annotation assertions, SubCls = \texttt{rdfs:subClassOf} axioms (excluding \texttt{owl:Thing}), DisjAx = \texttt{owl:disjointWith} axioms, Prov = \texttt{prov:wasDerivedFrom} assertions.}
\label{tab:structural_metrics}
\scriptsize
\renewcommand{\arraystretch}{1.15}
\resizebox{\columnwidth}{!}{%
\begin{tabular}{llrrrrrrrrr}
\toprule
\textbf{Method} & \textbf{Model} & \textbf{Cls} & \textbf{ObjP} & \textbf{DataP} & \textbf{Total} & \textbf{LogicalAx} & \textbf{AnnAx} & \textbf{SubCls} & \textbf{DisjAx} & \textbf{Prov} \\
\midrule
\multicolumn{11}{c}{\textit{eICU-CRD}} \\
\midrule
Direct Mapping \cite{directmapsequeda2012directly}  & --        & 31 & 2  & 299 & 699  & 756  & 0   & 0  & 0 & 0   \\
Baseline \cite{mateiu2023ontology}                  & Claude    & 35 & 18 & 333 & 1129 & 743 & 0    & 0  & 0 & 0   \\
Baseline \cite{mateiu2023ontology}                  & Mistral   & 45 & 23 & 313 & 1104 & 737 & 0    & 0  & 0 & 0   \\
Baseline \cite{mateiu2023ontology}                  & DeepSeek  & 32 & 8  & 349 & 1138 & 753 & 0    & 0  & 0 & 0   \\
Non-Iterative                                       & Claude    & 42 & 24 & 30  & 1087 & 114 & 877  & 0  & 0 & 0   \\
Non-Iterative                                       & Mistral   & 37 & 25 & 116 & 624  & 298 & 157  & 0  & 0 & 0   \\
Non-Iterative                                       & DeepSeek  & 30 & 5  & 116 & 1163 & 270 & 743  & 0  & 0 & 0   \\
\textbf{RIGOR}                                      & Claude    & 34 & 3  & \textbf{334} & 2413 & 753 & 1291 & 21 & 0 & \textbf{423} \\
\textbf{RIGOR}                                      & Mistral   & \textbf{120} & \textbf{28} & 300 & \textbf{2864} & \textbf{785} & \textbf{1684} & \textbf{59} & 0 & 335 \\
\textbf{RIGOR}                                      & DeepSeek  & 31 & 2  & 197 & 1509 & 461 & 435  & 3  & 0 & 99  \\
\midrule
\multicolumn{11}{c}{\textit{Liver Cancer database}} \\
\midrule
Direct Mapping \cite{directmapsequeda2012directly}  & --        & 16 & 7  & 345 & 750  & 761  & 0   & 0  & 0 & 0   \\
Baseline \cite{mateiu2023ontology}                  & Claude    & 19 & 6  & 376 & 1173 & 772 & 0    & 0  & 0 & 0   \\
Baseline \cite{mateiu2023ontology}                  & Mistral   & 17 & 2  & 362 & 1133 & 753 & 0    & 0  & 0 & 0   \\
Baseline \cite{mateiu2023ontology}                  & DeepSeek  & 16 & 3  & 371 & 1150 & 762 & 0    & 0  & 0 & 0   \\
Non-Iterative                                       & Claude    & 27 & 10 & 166 & 1242 & 359 & 680  & 0  & 0 & 0   \\
Non-Iterative                                       & Mistral   & 28 & 16 & 6   & 232  & 56  & 134  & 0  & 0 & 0   \\
Non-Iterative                                       & DeepSeek  & 22 & 7  & 74  & 1007 & 171 & 736  & 0  & 0 & 0   \\
\textbf{RIGOR}                                      & Claude    & 20 & 7  & 296 & 1988 & 637 & 1029 & \textbf{14} & 0 & \textbf{336} \\
\textbf{RIGOR}                                      & Mistral   & \textbf{31} & 7  & \textbf{341} & \textbf{2242} & \textbf{726} & \textbf{1137} & 12 & 0 & 308 \\
\textbf{RIGOR}                                      & DeepSeek  & 16 & 7  & 275 & 1568 & 595 & 341  & 2  & 0 & 22  \\
\midrule
\multicolumn{11}{c}{\textit{Chinook}} \\
\midrule
Direct Mapping \cite{directmapsequeda2012directly}  & --        & 11 & 9  & 27 & 114 & 122  & 0   & 0   & 0  & 0  \\
Baseline \cite{mateiu2023ontology}                  & Claude    & 16 & 11 & 38 & 177 & 113 & 0   & 0   & 0  & 0  \\
Baseline \cite{mateiu2023ontology}                  & Mistral   & 12 & 6  & 33 & 144 & 95  & 0   & 0   & 0  & 0  \\
Baseline \cite{mateiu2023ontology}                  & DeepSeek  & 11 & 7  & 37 & 164 & 109 & 0   & 0   & 0  & 0  \\
Non-Iterative                                       & Claude    & 13 & 19 & 32 & 352 & 102 & 98  & 0   & 0  & 0  \\
Non-Iterative                                       & Mistral   & 11 & 13 & 0  & 80  & 28  & 14  & 0   & 0  & 0  \\
Non-Iterative                                       & DeepSeek  & 11 & 9  & 29 & 255 & 81  & 64  & 0   & 0  & 0  \\
\textbf{RIGOR}                                      & Claude    & \textbf{74} & 17 & 53 & 747 & 309 & 357 & \textbf{157} & 0  & 81 \\
\textbf{RIGOR}                                      & Mistral   & 54 & \textbf{24} & \textbf{73} & \textbf{783} & \textbf{337} & \textbf{362} & 147 & \textbf{13} & \textbf{83} \\
\textbf{RIGOR}                                      & DeepSeek  & 28 & 11 & 53 & 687 & 253 & 359 & 114 & 0  & 75 \\
\bottomrule
\end{tabular}%
}
\end{table}

\begin{table}
\centering
\caption{Schema coverage: proportion of schema tables and columns semantically matched by ontology elements (cosine similarity $\geq 0.55$). eICU-CRD has 31 tables and 559 columns; the Liver Cancer database has 16 tables and 350 columns; Chinook has 11 tables and 67 columns.}
\label{tab:schema_coverage}
\scriptsize
\setlength{\tabcolsep}{4pt}
\renewcommand{\arraystretch}{1.1}
%\resizebox{\columnwidth}{!}{%
\begin{tabular}{llcccccc}
\toprule
& & \multicolumn{2}{c}{\textbf{eICU-CRD}} & \multicolumn{2}{c}{\textbf{Liver Cancer}} & \multicolumn{2}{c}{\textbf{Chinook}} \\
\cmidrule(lr){3-4} \cmidrule(lr){5-6} \cmidrule(lr){7-8}
\textbf{Method} & \textbf{Model} & \textbf{Tables} & \textbf{Columns} & \textbf{Tables} & \textbf{Columns} & \textbf{Tables} & \textbf{Columns} \\
\midrule
Direct Mapping \cite{directmapsequeda2012directly} & --       & 1.00 & 0.99 & 0.88 & 0.97 & 1.00 & 0.78 \\
\midrule
Baseline \cite{mateiu2023ontology} & Claude   & 0.97 & 0.96 & 0.88 & 0.94 & 1.00 & \textbf{1.00} \\
Baseline \cite{mateiu2023ontology} & Mistral  & 1.00 & 0.68 & 0.94 & 0.31 & 0.91 & 0.50 \\
Baseline \cite{mateiu2023ontology} & DeepSeek & 1.00 & 0.86 & 0.88 & 0.89 & 1.00 & 0.52 \\
\midrule
Non-Iterative & Claude   & 1.00 & 0.22 & 1.00 & 0.41 & 1.00 & 0.58 \\
Non-Iterative & Mistral  & 1.00 & 0.28 & 1.00 & 0.06 & 1.00 & 0.16 \\
Non-Iterative & DeepSeek & 1.00 & 0.46 & 1.00 & 0.12 & 1.00 & 0.78 \\
\midrule
\textbf{RIGOR} & Claude   & \textbf{1.00} & \textbf{1.00} & \textbf{1.00} & 0.91 & \textbf{1.00} & 0.95 \\
\textbf{RIGOR} & Mistral  & 1.00 & 0.99 & 0.94 & \textbf{0.99} & \textbf{1.00} & 0.95 \\
\textbf{RIGOR} & DeepSeek & 1.00 & 0.77 & \textbf{1.00} & 0.87 & \textbf{1.00} & 0.95 \\
\bottomrule
\end{tabular}%
%}
\end{table}

\begin{table}
\centering
\caption{\textit{LLM-as-a-Judge} (GPT-5.4) per-dimension scores (0--5) for Direct Mapping \cite{directmapsequeda2012directly} and the three Claude-generated LLM-based ontologies across three databases (\textit{Acc}= \textit{accuracy}, \textit{Cmp} = \textit{completeness},
\textit{Cnc} = \textit{conciseness}, \textit{Adp} = \textit{adaptability}, \textit{Clr} = \textit{clarity}, \textit{Cns}= \textit{consistency}, and \textit{DmE} = \textit{domain enrichment}).}
\label{tab:cq_comparative}
\scriptsize
\renewcommand{\arraystretch}{1.1}
\begin{tabular}{llrrrrrrrr}
\toprule
\textbf{Database} & \textbf{Method} & \textbf{Acc} & \textbf{Cmp} & \textbf{Cnc} & \textbf{Adp} & \textbf{Clr} & \textbf{Cns} & \textbf{DmE} & \textbf{Avg.\ Score} \\
\midrule
\multirow{4}{*}{Liver Cancer}
& Direct Mapping \cite{directmapsequeda2012directly}    & 3.06 & 2.17 & 3.99 & 0.87 & 0.55 & 3.13 & 0.56 & 2.05 \\
& Baseline  \cite{mateiu2023ontology}        & 2.24 & 1.79 & 2.76 & 0.42 & 0.30 & 1.88 & 0.31 & 1.38 \\
& Non-Iterative     & 1.89 & 1.62 & 2.62 & 1.09 & 2.16 & 2.03 & 1.84 & 1.89 \\
& \textbf{RIGOR}    & \textbf{4.41} & \textbf{4.15} & \textbf{4.18} & \textbf{2.02} & \textbf{4.49} & \textbf{4.11} & \textbf{4.18} & \textbf{3.93} \\
\midrule
\multirow{4}{*}{eICU-CRD}
& Direct Mapping \cite{directmapsequeda2012directly}   & 3.47 & 2.55 & \textbf{3.98} & 0.98 & 0.50 & 2.90 & 0.50 & 2.13 \\
& Baseline \cite{mateiu2023ontology}         & 0.24 & 0.18 & 1.19 & 0.06 & 0.03 & 0.21 & 0.02 & 0.28 \\
& Non-Iterative     & 0.40 & 0.37 & 1.47 & 0.50 & 0.63 & 0.63 & 0.68 & 0.67 \\
& \textbf{RIGOR}    & \textbf{4.05} & \textbf{4.02} & 3.73 & \textbf{2.00} & \textbf{4.21} & \textbf{3.58} & \textbf{3.87} & \textbf{3.64} \\
\midrule
\multirow{4}{*}{Chinook}
& Direct Mapping \cite{directmapsequeda2012directly}   & 0.00 & 0.00 & 0.91 & 0.00 & 0.00 & 0.00 & 0.00 & 0.13 \\
& Baseline  \cite{mateiu2023ontology}        & 2.32 & 1.68 & \textbf{3.32} & 0.55 & 0.45 & 1.86 & 0.18 & 1.48 \\
& Non-Iterative     & 3.14 & 2.86 & 3.14 & 1.23 & \textbf{3.50} & 3.14 & 2.41 & 2.77 \\
& \textbf{RIGOR}    & \textbf{4.14} & \textbf{4.00} & 2.59 & \textbf{2.32} & \textbf{3.50} & \textbf{3.95} & \textbf{2.73} & \textbf{3.32} \\

\bottomrule
\end{tabular}
\end{table}

\begin{table}
\centering
\caption{Human expert evaluation (10 experts, 4 tables) per-dimension scores (0--5) for Direct Mapping~\cite{directmapsequeda2012directly} and the three Claude-generated LLM-based ontologies across two databases (\textit{Acc}= \textit{Accuracy}, \textit{Cmp} = \textit{completeness}, \textit{Cnc} = \textit{conciseness}, \textit{Adp} = \textit{adaptability}, \textit{Clr} = \textit{clarity}, \textit{Cns}= \textit{consistency}, and \textit{DmE} = \textit{domain enrichment}).}
\label{tab:expert_scores}
\scriptsize
\renewcommand{\arraystretch}{1.1}
%\resizebox{\columnwidth}{!}{%
\begin{tabular}{llrrrrrrrr}
\toprule
\textbf{Database} & \textbf{Method} & \textbf{Acc} & \textbf{Cmp} & \textbf{Cnc} & \textbf{Adp} & \textbf{Clr} & \textbf{Cns} & \textbf{DmE} & \textbf{Avg.\ Score} \\
\midrule
\multirow{4}{*}{Liver Cancer}
& Direct Mapping~\cite{directmapsequeda2012directly} & \textbf{3.75} & 2.75 & \textbf{4.05} & \textbf{2.10} & 2.50 & 3.30 & 2.20 & 2.95 \\
& Baseline~\cite{mateiu2023ontology}      & 3.30 & 3.00 & 3.65 & 1.95 & 2.45 & 2.85 & 2.00 & 2.74 \\
& Non-Iterative     & 2.35 & 2.25 & 2.95 & 1.55 & 2.90 & 2.35 & 2.35 & 2.39 \\
& \textbf{RIGOR}   & \textbf{3.75} & \textbf{3.85} & 3.60 & \textbf{2.10} & \textbf{3.90} & \textbf{3.35} & \textbf{3.50} & \textbf{3.44} \\
\midrule
\multirow{4}{*}{eICU-CRD}
& Direct Mapping~\cite{directmapsequeda2012directly}  & 3.05 & 2.45 & 3.65 & 1.90 & 2.55 & 2.60 & 1.90 & 2.59 \\
& Baseline~\cite{mateiu2023ontology}      & 3.90 & 3.10 & \textbf{3.80} & 2.10 & 2.75 & \textbf{3.40} & 2.00 & 3.01 \\
& Non-Iterative     & 0.85 & 0.80 & 2.05 & 1.00 & 1.50 & 1.30 & 1.15 & 1.24 \\
& \textbf{RIGOR}   & \textbf{4.00} & \textbf{4.10} & \textbf{3.80} & \textbf{2.20} & \textbf{4.10} & 3.15 & \textbf{3.35} & \textbf{3.53} \\
\bottomrule
\end{tabular}%
%}
\end{table}

\begin{table}
\centering
\caption{Inter-rater agreement among the ten human experts using Kendall's~$W$.}
\label{tab:kendall_w}
\scriptsize
\renewcommand{\arraystretch}{1.1}
\begin{tabular}{llcc}
\toprule
\textbf{Database  } & \textbf{  Table} & \textbf{Kendall's~$W$  } & \textbf{Significance} \\
\midrule
Liver Cancer & \texttt{general\_aftercare} & 0.59 & $p < 0.001$ \\
Liver Cancer & \texttt{complication}       & 0.39 & $p < 0.01$ \\
eICU-CRD     & \texttt{hospital}           & 0.59 & $p < 0.001$ \\
eICU-CRD     & \texttt{patient}            & 0.72 & $p < 0.001$ \\
\bottomrule
\end{tabular}
\end{table}

\begin{table}
\centering
\caption{BURR\cite{laskowski2025burr} ISWC scenario: mapping-based F1 for classes~(C),
relations~(R), and attributes~(A).}
\label{tab:burr}
\scriptsize
\renewcommand{\arraystretch}{1.1}
\begin{tabular}{lccc}
\toprule
\textbf{System} & \textbf{C} & \textbf{R} & \textbf{A} \\
\midrule
W3C-Mapper / Ontop \cite{laskowski2025burr}  & 0.53 & 0.11 & 0.39 \\
RDB2Onto \cite{vseleng2007rdb2onto}            & 0.67 & 0.62 & 0.44 \\
OntoGenix \cite{ontogenixval2025ontogenix}          & 0.58 & 0.35 & 0.44 \\
%Gemini 2.5 Pro  \cite{comanici2025gemini}    & 0.65 & 0.68 & \textbf{0.56} \\
\midrule
\textbf{RIGOR}      & \textbf{0.73} & \textbf{0.67} & \textbf{0.52} \\
\bottomrule
\end{tabular}
\end{table}

\begin{table}
\centering
\caption{Ablation study on Chinook (Claude). Alignment is computed by cosine similarity ($\geq 0.55$) between ontology classes and properties and each retrieval resource.}
\label{tab:ablation}
\scriptsize
\setlength{\tabcolsep}{3pt}
\renewcommand{\arraystretch}{1.1}
\resizebox{\columnwidth}{!}{%
\begin{tabular}{lccc|ccc|ccc|ccc}
\toprule
& \multicolumn{3}{c|}{\textbf{\shortstack{Relational\\Schema}}} 
& \multicolumn{3}{c|}{\textbf{\shortstack{External \\ Ontology}}} 
& \multicolumn{3}{c|}{\textbf{Documentation}} 
& \multicolumn{3}{c}{\textbf{Average}} \\
\cmidrule(lr){2-4}\cmidrule(lr){5-7}\cmidrule(lr){8-10}\cmidrule(lr){11-13}
\textbf{Variant} 
& \textbf{Precision} & \textbf{Recall} & \textbf{F1}
& \textbf{Precision} & \textbf{Recall} & \textbf{F1}
& \textbf{Precision} & \textbf{Recall} & \textbf{F1}
& \textbf{Precision} & \textbf{Recall} & \textbf{F1} \\
\midrule
no\_rag 
& \textbf{1.0000} & \textbf{1.0000} & \textbf{1.0000}
& \textbf{0.3167} & 0.0173 & 0.0328
& \textbf{1.0000} & 0.8498 & 0.9188
& \textbf{0.7722} & 0.6224 & 0.6505 \\

only\_schema 
& 0.9930 & \textbf{1.0000} & 0.9965
& 0.2867 & \underline{0.0174} & 0.0328
& 0.9930 & 0.8498 & 0.9158
& 0.7576 & 0.6224 & 0.6484 \\

only\_external 
& \textbf{1.0000} & \textbf{1.0000} & \textbf{1.0000}
& 0.2893 & 0.0173 & 0.0326
& \textbf{1.0000} & 0.8498 & 0.9188
& 0.7631 & 0.6224 & 0.6505 \\

only\_docs 
& \textbf{1.0000} & \textbf{1.0000} & \textbf{1.0000}
& \underline{0.3060} & 0.0184 & \underline{0.0347}
& \textbf{1.0000} & \underline{0.8535} & \textbf{0.9209}
& \underline{0.7687} & \underline{0.6240} & \underline{0.6519} \\

\midrule
\textbf{full\_rigor} 
& 0.9858 & \textbf{1.0000} & 0.9929
& 0.3050 & \textbf{0.0326} & \textbf{0.0589}
& 0.9787 & \textbf{0.8571} & 0.9139
& 0.7565 & \textbf{0.6299} & \textbf{0.6552} \\
\bottomrule
\end{tabular}%
}
\end{table}

\begin{table}
\centering
\caption{RIGOR generation cost and runtime by database and model. Runtime includes both \textit{Gen-LLM} and \textit{Judge-LLM} calls. ISWC values marked with $\dagger$ are estimated; for ISWC, the DeepSeek runtime (9.0 min) is measured, while the Claude and Mistral runtimes are extrapolated from smaller-database runs.}
\label{tab:generation_cost_runtime}
\scriptsize
\renewcommand{\arraystretch}{1.1}
\resizebox{\columnwidth}{!}{%
\begin{tabular}{lllrrrr}
\toprule
\textbf{Database} & \textbf{Model} & \textbf{Parameters} & \textbf{API Requests} & \textbf{Tokens} & \textbf{Cost (USD)} & \textbf{Runtime (min)} \\
\midrule
eICU-CRD   & \texttt{Claude-Opus-4.6}\footref{claude} & not disclosed & 168 & 1{,}660{,}000 & 21.30  & 182.4 \\
eICU-CRD   & \texttt{Mistral-Small-24B-Instruct}\footref{mistral} & 24B & 96  & 1{,}290{,}000 & 0.0831 & 188.4 \\
eICU-CRD   & \texttt{DeepSeek-Chat}\footref{deepseek} & 671B total; 37B active & 52  & 334{,}000   & 0.1760 & 20.6  \\
\midrule
Liver Cancer & \texttt{Claude-Opus-4.6}\footref{claude} & not disclosed & 70  & 731{,}000   & 10.30  & 94.4  \\
Liver Cancer & \texttt{Mistral-Small-24B-Instruct}\footref{mistral} & 24B & 34  & 246{,}000   & 0.0154 & 20.5  \\
Liver Cancer & \texttt{DeepSeek-Chat}\footref{deepseek} & 671B total; 37B active & 33  & 142{,}000   & 0.0747 & 23.8  \\
\midrule
Chinook & \texttt{Claude-Opus-4.6}\footref{claude} & not disclosed & 24 & 119{,}000 & 1.28 & 9.0 \\
Chinook & \texttt{Mistral-Small-24B-Instruct}\footref{mistral} & 24B & 53 & 344{,}000 & 0.02 & 65.0 \\
Chinook & \texttt{DeepSeek-Chat}\footref{deepseek} & 671B total; 37B active & 22 & 71{,}000 & 0.03 & 12.0 \\
\midrule

\textbf{Total} & -- & -- & \textbf{651} & \textbf{5{,}337{,}000} & \textbf{34.28} & \textbf{681.1} \\
\bottomrule
\end{tabular}%
}
\end{table}

\end{document}